\documentclass{aa}
\usepackage{txfonts}
\usepackage{graphicx}

\def\H2{H$_2$}

\begin{document}

\title{SOLIS. XIX. The chemically rich SVS13-B protostellar jet}

 
\author{C. Codella \inst{1,2}
\and
E. Bianchi \inst{3,1} 
\and
L. Podio \inst{1}
\and
M. De Simone \inst{4,1}
\and
A. L\'opez-Sepulcre \inst{5,2}
\and
C. Ceccarelli \inst{2}
\and
P. Caselli \inst{6}
}

\institute{
INAF, Osservatorio Astrofisico di Arcetri, Largo E. Fermi 5,
50125 Firenze, Italy
\and
Univ. Grenoble Alpes, CNRS, Institut de
Plan\'etologie et d'Astrophysique de Grenoble (IPAG), 38000 Grenoble, France
\and
Excellence Cluster ORIGINS, Boltzmannstraße 2, 85748, Garching bei Mu\"unchen, Germany
\and
ESO, Karl Schwarzchild Srt. 2, 85478 Garching bei M\"unchen, Germany
\and
Institut de Radioastronomie Millim\'etrique, 300 rue de la Piscine, Domaine
Universitaire de Grenoble, 38406, Saint-Martin d'H\`eres, France
\and
Center for Astrochemical Studies, Max-Planck-Institut f\"{u}r extraterrestrische Physik (MPE), Gie$\beta$enbachstr. 1, D-85741 Garching, Germany
}

\offprints{C. Codella, \email{claudio.codella@inaf.it}}
\date{Received date; accepted date}

\authorrunning{Codella et al.}
\titlerunning{Molecules in the protostellar SVS13-B outflow}

\abstract 
%
{As part of the IRAM Large Program SOLIS, we imaged the SVS13 protostellar system in line emission to kinematically separate jets from the large-scale outflows and static envelopes and to investigate their chemistry. 
%
Using the NOEMA interferometer, we imaged the protostellar sources SVS13-A and SVS13-B in SiO(2--1),  SO(2$_{\rm 3}$--1$_{\rm 2}$),
CS(2--1), and CH$_3$OH(2$_{\rm k,k}$--1$_{\rm k,k}$) at a spatial resolution of 
$\sim$ 2$\arcsec$--3$\arcsec$ (600--900 au). 
An SiO(2--1) image with a smaller beam (1$\farcs$5, i.e. 450 au) was also produced. 
%
The CS and SO emission traces the $\sim 5000$ au envelope that hosts the SVS13-A and VLA3 young stellar objects,  
and CH$_3$OH probes the compact hot corino associated with SVS13-A (T$\sim 100-110$ K). In addition, CS blue-shifted emission reveals a molecular  shell in the direction of the jet driven by SVS13-A that is revealed by high-velocity SiO, SO and low-velocity H$_2$ emission (PA $\sim 155\degr$).

We also imaged the protostellar jet driven by SVS13-B in SiO, and in SO, CS, and CH$_3$OH for the first time as well, along PA$\sim167\degr$. The molecules peak at different distances from the driving source: SiO(2--1) peaks at $\sim$ 1600 au, and SO(2$_{\rm 3}$--1$_{\rm 2}$), CS(2--1) and CH$_3$OH(2$_{\rm k,k}$--1$_{\rm k,k}$) peak at $\sim$ 2000--2850 au.
Moreover, SiO(2--1) emits at larger distances than SiO(5--4), indicating a lower excitation at a larger distance from the protostar.
The observed species also show different velocity distributions: SiO peaks at velocities up to $+35$ km s$^{-1}$ (red) and $-20$ km s$^{-1}$ (blue) with respect to the systemic velocity, SO and CS peak at $\pm$ 10 km s$^{-1}$, and CH$_3$OH is at low velocities of $\pm$ 4 km s$^{-1}$.
%
The multi-species observations revealed a
stratified chemical structure in the jet of SVS13-B. A jet-like component with a transversal size $\leq$ 450 au is traced by SiO, which is efficiently formed in
high-velocity shocks ($>25$ km s$^{-1}$) by  sputtering and vaporisation of the grain cores and mantles.
A slower and wider (transversal size $\sim$ 750 au) component is probed by methanol, which is released from dust mantles at lower shock velocities ($<10$ km s$^{-1}$).
The SO and CS emission traces an intermediate component with respect to the components probed by SiO and CH$_3$OH.
High spatial resolution imaging (down to 10 au) of the jet of SVS13-B in multiple species will aid in reconstructing the chemistry of shocked material in protostellar jets.
}


\maketitle

\section{Introduction}\label{sec:intro}

Stars similar to our Sun originate from high-density 
($\ge$ 10$^{5}$ cm$^{-3}$) cores that are distributed inside filaments. At the so-called Class 0 stage, a protostar with an age of approximately 10$^4$ years \citep[e.g., ][]{Andre2000} accretes mass at an high rate from its disk \citep[e.g., ][]{Pineda2023}.
To enable accretion from the disk onto the protostar, angular momentum needs to be removed from the disk by jets, outflows, and disk winds \citep{Shu1987}.
Therefore, the ejection process is a natural outcome of the star formation process. 

Supersonic ($\sim$ 100 km s$^{-1}$) collimated jets
are ejected from the inner region of the star-disk system perpendicular to the accretion disks, and
they accelerate the dense material of the cloud that surrounds the protostar. This creates slower ($\sim$ 10 km s$^{-1}$)
molecular outflows that can be observed up to large scales
(fractions of a parsec) mainly through CO \citep[e.g., ][and references therein]{Lada1985,Frank2014}. 
Jets driven by Class 0 sources are detected using molecular species, for instance, SiO, the classical tracer, and CO emission at high velocities at (sub)millimeter wavelengths \citep[e.g., ][and references therein]{Lee2020,Podio2016,Podio2021},
and by H$_2$ in the infrared spectral window, as shown by the recent spectacular images by the James Webb Space Telescope (JWST)
\citep[e.g., ][]{ray2023,caratti2024}.

The first detections of high-velocity bullets that may probe the protostellar jets in other molecular tracers were reported by \citet{Tafalla2010}, who observed the outflows of L1448-mm and IRAS 04166+2706 owith the IRAM 30 m antenna. Subsequently, interferometric studies, such as the statistical study by \citet{Podio2021} as well as a number of other works, confirmed that SO is detected not only in the terminal shocks, but can also be used to trace high-velocity collimated jets, similarly to SiO \citep[e.g., ][]{Lee2007,Codella2014,Podio2015,Santangelo2015}.
SiO jets are expected to rotate \citep[see the HH212 case by][]{Lee2017c}, and the detection of slower molecular gas that rotates at wider angles for instance in SO and SO$_2$ supports the assumption that extended disk winds occur around the jets, and that they originate from a wider disk region \citep[e.g., ][]{Tabone2017,Lee2018}.

The advent of the Atacama Large Millimeter Array interferometer\footnote{https://www.almaobservatory.org} (ALMA), with its unique combination of a high spatial resolution and high sensitivity in the (sub)mm spectral window allowed us to observe other species associated with the mass loss from Sun-like protostars. 
More specifically, the ALMA observational campaigns of the isolated 
HH 212 protostar, which reached a spatial resolution of $\sim$ 10 au, showed that
emission due to S-bearing species such as SO and SO$_2$ are valuable 
tools for revealing the magnetohydrodynamic disk wind, which is slower ($\sim$ 30 km s$^{-1}$) than the collimated SiO jet \citep{Podio2015,Tabone2017,Tabone2020,Lee2018}.
In addition, methanol (CH$_3$OH) has also very recently been proposed to trace the base of the disk wind in L1448 and NGC1333-IRAS4A \citep{Nazari2024,DeSimone2024}.
The chemical richness of molecular jets was also studied for a few targets in Serpens, and emission in CO, SiO, H$_2$CO, and HCN was reported \citep{Tec2019}.  
These recent studies showed that it is timely to investigate the chemical composition of the molecular jet, and that this should be extended to various targets, in particular, isolated protostars, to effectively distinguish the origin of the emission from different species and at different velocities.

In the context of the 
the IRAM/NOEMA interferometer\footnote{\url{http://www.iram-institute.org/}} Large Program SOLIS\footnote{\url{http://solis.osug.fr/}} \citep[Seeds Of Life In Space:][]{Ceccarelli2017}, we observed the protostellar cluster SVS13 in Perseus (see Sect. \ref{sec:svs13}) in emission lines of SiO, SO, CS, and CH$_3$OH (see Sect. \ref{sec:observations}). In this paper, we present the results from the analysis of the continuum and line emission maps (Sects. \ref{sec:results-continuum} and \ref{sec:results-line}). We unveil the different gas components associated with the protostellar sources, namely the envelope and hot corino, the molecular shell associated with SVS13-A (Sects. \ref{sec:envelope}, and \ref{sec:shells}), and the protostellar jet driven by SVS13-B (Sect. \ref{sec:jetB}). In Sect. \ref{sec:discussion} we discuss our results, for which we focus on the chemical composition of the SVS13-B jet (Sects. \ref{sec:differentregions}, \ref{sec:excitation}) and of the shell that is driven by SVS13-A (Sect. \ref{sec:cs-so}). Finally, our conclusions are summarised in Sect. \ref{sec:conclusions}.

\section{The SVS13 protostellar cluster} 
\label{sec:svs13}

The SVS13 protostellar system is located in the well-known NGC1333 cluster in the Perseus region at a distance of 299 $\pm$ 14 pc \citep{Zucker2018}.
It contains several protostellar solar analogues that are classified as Class 0 ($\geq$ 10$^4$ yr; SVS13-B, SVS13-C) or Class I ($\sim$ 10$^5$; SVS13-A). An additional protostar, VLA3, is located close to SVS13-A, and its evolutionary stage has not yet been firmly assessed \citep[see e.g., ][]{Chini1997,Maury2019}.
SVS13-A is a close binary source composed by VLA4A and VLA4B with a separation of 0$\farcs$3 (90 au) 
\citep{Anglada2000}. The region was extensively observed
at different wavelengths in the past decades \citep[see e.g.][and references therein]{Chini1997, Looney2000, Chen2009, Tobin2016, Lefloch2018, Ceccarelli2017, Maury2019, Maret2020, Diaz2022, Codella2021, Bianchi2022a, Bianchi2022b, Bianchi2023, Hsieh2023, Hsieh2024}.
Subsequently, the mass loss processes from the protostars were analysed as well. The SVS13A binary system drives an extended molecular outflow that has been detected since the 1990s using single-dish observations and the standard CO and SiO outflow/jet tracers \citep[e.g.
][]{Bachiller1990,Bachiller1998,Lefloch1998a,Codella1999}.
SVS13-A is also associated with the well-known HH7–11 system, 
which is a chain of bright Herbig-Haro (HH) knots about 1$\arcmin$ long \citep{Reipurth1993}. At low angular resolution \citep[larger than 10$\arcsec$, e.g.][]{Bachiller1990}, the HH7–11 chain is distributed along the blue-shifted lobe of the CO outflow. 
Further single-dish observations \citep{Lefloch1998a,Codella1999}
revealed that SiO lies in a structure that seems complementary to that
of the HH system.
Successively, high-spatial resolution imaging in the infrared and millimeter
spectral range \citep{Bachiller1998,Hodapp2014,Lefevre2017,Tobin2018} uncovered a complex scenario. More precisely, at least two jets are launched from the SVS13-A binary
\citep[in the SE direction, see Fig. 5 by][]{Lefevre2017}: (i) one jet is
associated with CO/H$_2$ emission, and (ii) the other jet lies in the
direction of the HH7-11 chain.

For the SVS13-B protostellar jet, the mass loss activity was revealed by \citet{Bachiller1998}. The authors imaged an SiO(2--1) bipolar jet driven by SVS13-B with the red gas to the NW and the blue gas to the SE with the IRAM-PdBI array 
at 3.5mm. Although the velocity bandwidth was relatively limited ($\sim$ 60 km s$^{-1}$), the results by \citet{Bachiller1998} revealed a new region where to study the kinematics and chemistry of protostellar jets. \citet{Podio2021} further imaged the jet of SVS13-B 
using the IRAM PdBI at 1.3mm in CO(2--1) and SiO(5--4). Weak SO(5$_6$-4$_5$) emission was revealed by \citet{Podio2021}, 
while a faint SO(2$_3$-1$_2$) jet was reported by \citet{Codella2021}, but was not analysed because the paper focused on
SVS13-A. These findings call for further observations using different molecular species that trace shocked gas.

\section{Observations} \label{sec:observations}

 \begin{figure}
\begin{center}
\includegraphics[angle=0,scale=0.7]{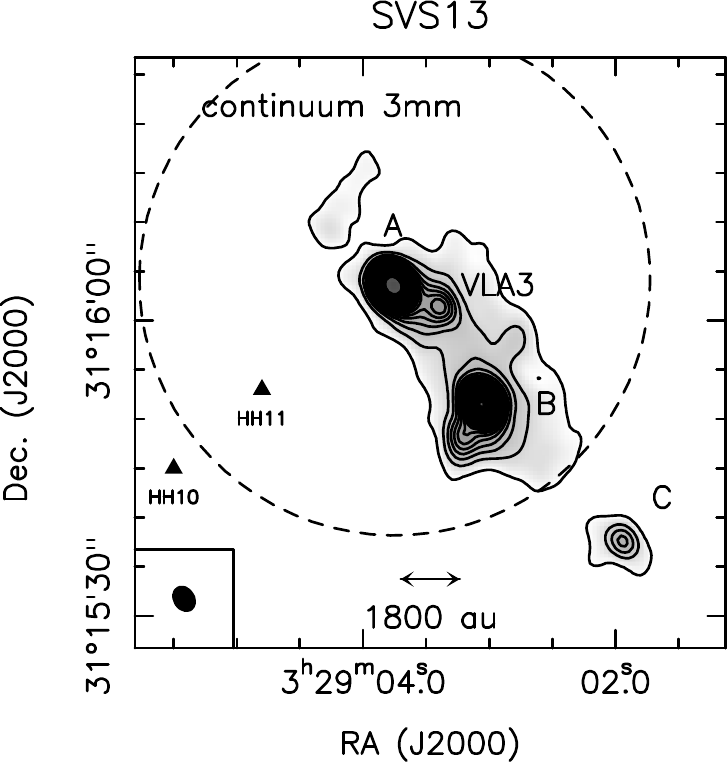}
\caption{Dust continuum emission at 3mm (greyscale and contours) of the SVS13 multiple system observed with NOEMA using configurations C and D
(setup 1; see Sect. \ref{sec:observations}). The first contours are at 3$\sigma$ (210 $\mu$Jy beam$^{-1}$), and the steps are in 10$\sigma$. The synthesised beam (bottom left corner) is 2$\farcs$81 $\times$ 2$\farcs$23 (PA = 31$\degr$). The brightness peaks reveal the positions of the SVS13-A, VLA3, SVS13-B, and SVS13-C protostars (see text). The dashed circle delimitates the FoV of the 3mm observations of 51$\arcsec$. The black triangles indicate the positions of the HH11 and HH10 Herbig-Haro objects.
\citep[e.g.][]{Bally1996}}.\label{fig:continuum}
\end{center}
\end{figure}

Our observations (see Tab. \ref{tab:lines}) 
of SVS13 were obtained at 3mm from
two complementary datasets as part of the IRAM/NOEMA  Large Program SOLIS.
The CS, SO, and CH$_3$OH observations (setup 1) were obtained using two tracks (configurations C and D) between 2016 and 2017, using eight antennas. The shortest and longest projected baselines are 22 m and 304 m, respectively. 
The field of view (FoV) was $\sim$ 50$\arcsec$, and the largest angular scale (LAS) is about 15$\arcsec$. 
The SiO emission (setup 2) was obtained with two tracks (configurations A and C) in 2018 using nine antennas. In this case, the baselines ranged from 64 m to 760 m, with a FoV = 57$\arcsec$ and a LAS = 5$\arcsec$.
For both setups, the images were centred at $\alpha_{\rm J2000}$ = 03$^{\rm h}$ 29$^{\rm m}$ 03$\fs$76, $\delta_{\rm J2000}$ = +31$\degr$ 16$\arcmin$ 03$\farcs$0 in order to observe SVS13-A, VLA3, and SVS13-B. 

In order to sample the CS(2--1), 
SO(2$_{\rm 3}$--1$_{\rm 2}$), and CH$_3$OH (2$_{\rm k,k}$--1$_{\rm k,k}$)
frequencies, the WideX backend was used for setup 1. It provided a bandwidth of 
$\sim$ 3.6 GHz (from 95.8 GHz to 99.4 GHz) with a spectral resolution of 2 MHz ($\sim$ 6 km s$^{-1}$). In addition, 320 MHz wide narrowband backends were used that provided a spectral resolution of around 0.5 km s$^{-1}$ .
The calibration was performed following the standard procedures using GILDAS-CLIC\footnote{\url{http://www.iram.fr/IRAMFR/GILDAS}}. 
The bandpass was calibrated on 3C84, the absolute flux was calibrated using LkH$\alpha$101, MWC249, and the phase was calibrated using 0333+321.
The absolute flux scale has an uncertainty of $\leqslant$10$\%$. 
The datacubes that included CS, SO, and CH$_3$OH  were continuum substracted in the $uv$ domain after line-free channels were identified, and they were subsequently imaged using natural weighting. They were restored with a clean beam of about 2$\farcs$6 $\times$ 2$\farcs$2 (see Table \ref{tab:lines}).
The clean beam of the continuum image is 2$\farcs$81 $\times$ 2$\farcs$23 (PA = 31$\degr$), with an rms noise of 70 $\mu$Jy beam$^{-1}$.
The line images were produced using natural weighting and were restored with a typical clean beam of 2$\farcs$7 $\times$ 2$\farcs$2 (Table \ref{tab:lines}). 

\begin{figure}
\begin{center}
\includegraphics[angle=0,scale=0.4]{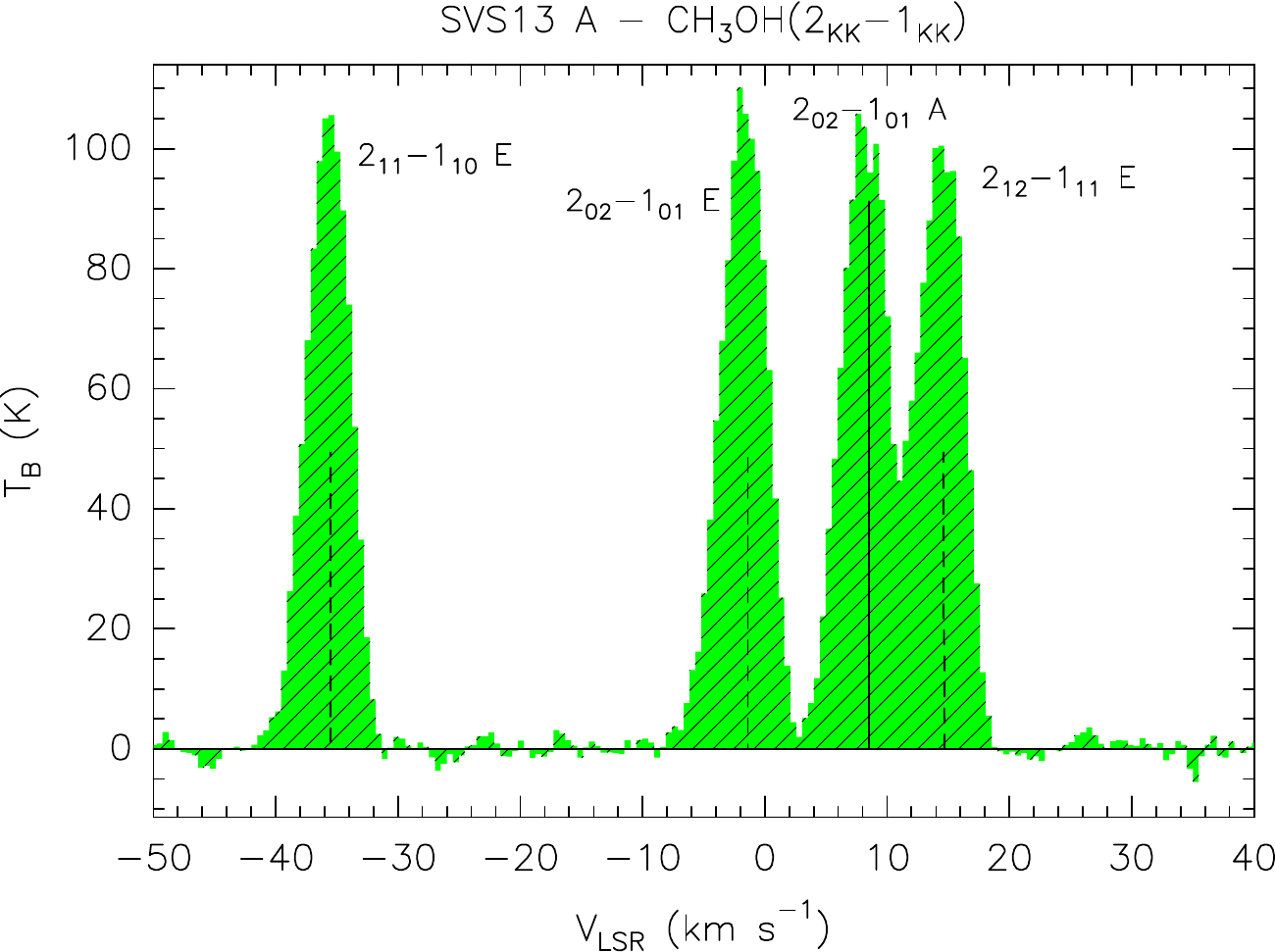}
\caption{CH$_3$OH(2$_{\rm kk}$--1$_{\rm kk}$) spectrum (brightness temperature, $T_{\rm B}$, scale) extracted at the continuum peak position of SVS13-A
(Tab. \ref{tab:continuum}). The spectra were multiplied by a factor 37, assuming an emitting region of 0$\farcs$4 (see Sect. \ref{sec:envelope}). 
The transitions producing the emission methanol profiles are labelled, and the corresponding frequencies are marked by vertical black segments (see Table \ref{tab:lines}). The CH$_3$OH spectra are centred at the frequency of the 2$_{\rm 0,2}$--1$_{\rm 0,1}$ A transition: 96741.38 MHz.
The vertical black line at +8.5 km s$^{-1}$ shows the systemic velocity \citep{Podio2021}.} \label{fig:methanol-hotcorino}
\end{center}
\end{figure}

\begin{figure*}
\begin{center}
\includegraphics[angle=0,scale=0.6]{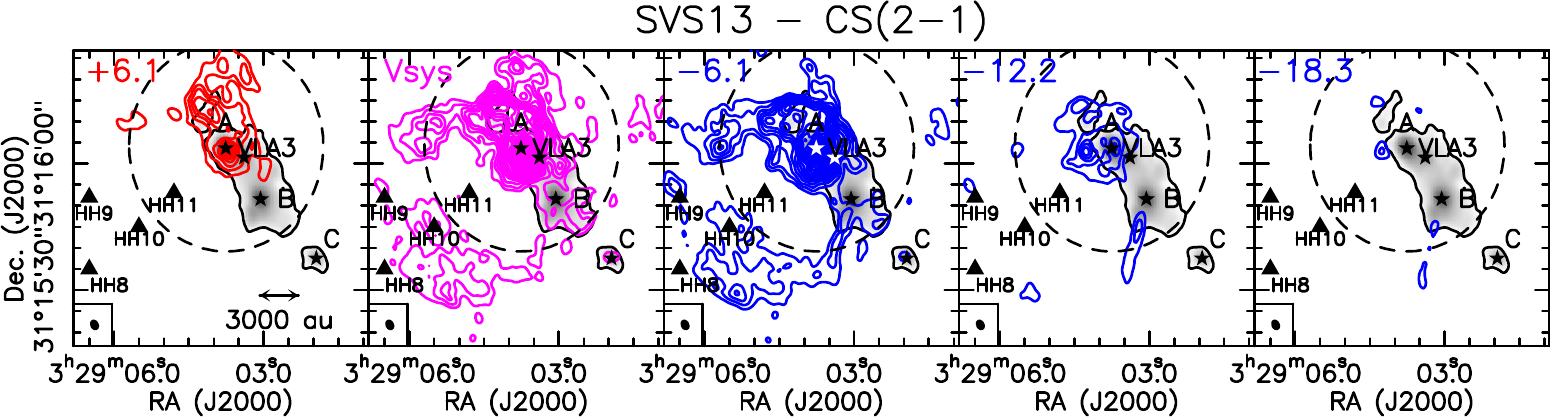}
\includegraphics[angle=0,scale=0.6]{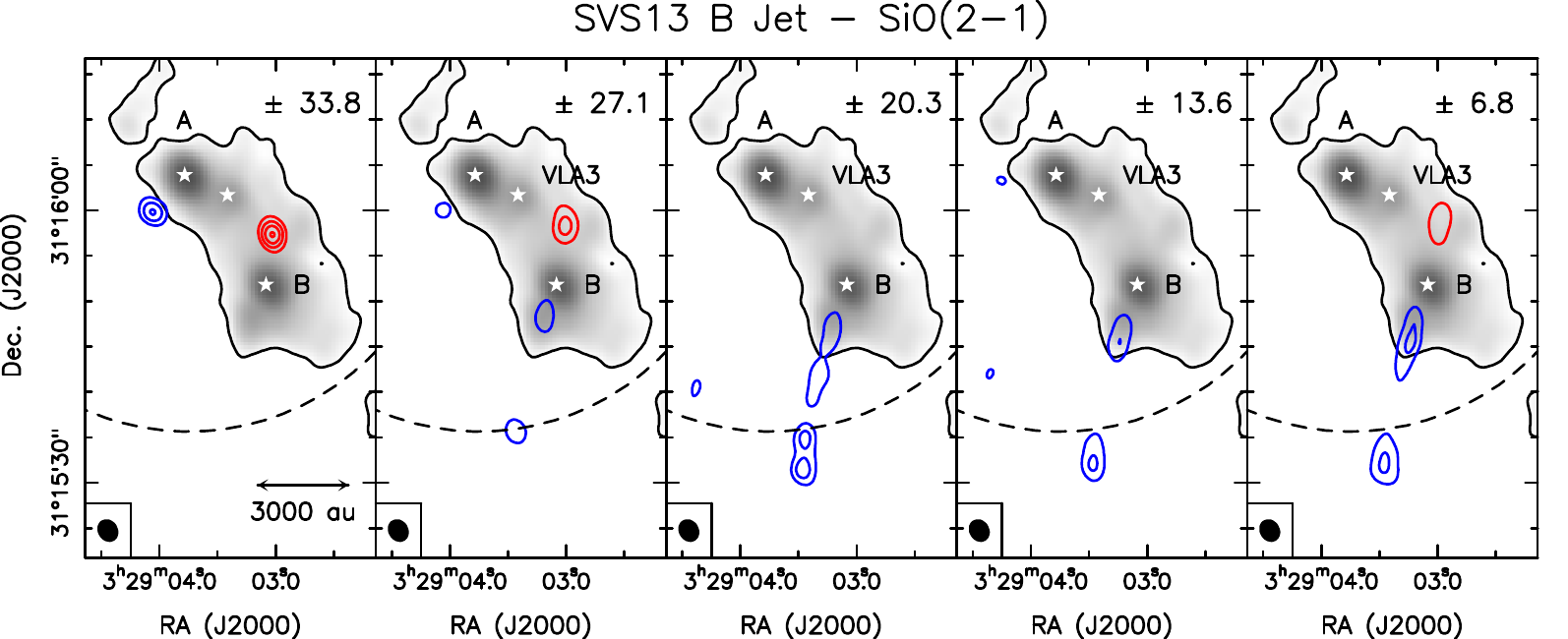}
\includegraphics[angle=0,scale=0.6]{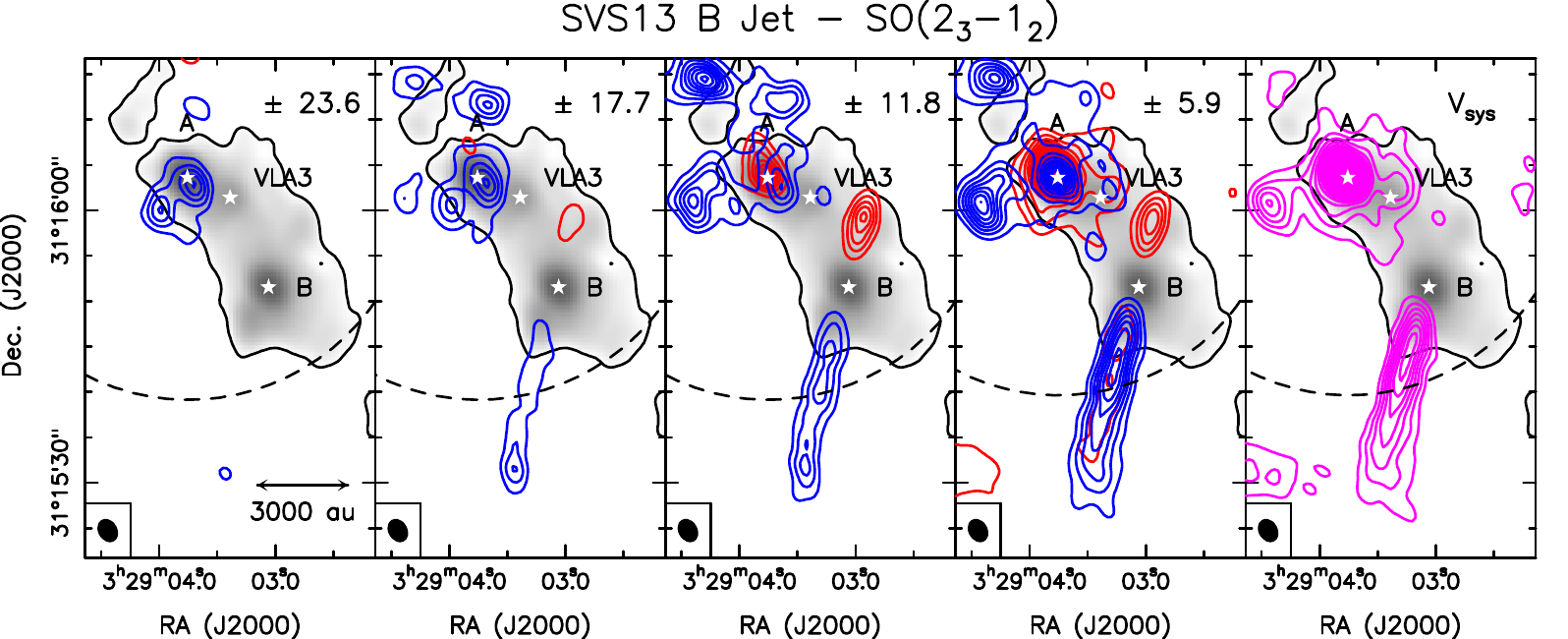}
\caption{Channel maps of red- and blue-shifted emission from CS(2--1), SiO(2--1), and SO(6$_6$--5$_5$) ({\it top, middle, and bottom panels}, respectively). Magenta shows the systemic velocity \citep[+8.5 km s$^{-1}$, ][]{Podio2021}. Each panel shows the emission shifted in velocity by the value given at the top. 
We show the 3mm continuum image (Fig. \ref{fig:continuum}) in greyscale (and black contours). The first contours and steps are 3$\sigma$ (3.6 mJy beam$^{-1}$, 2.4 mJy beam$^{-1}$, and 2.1 mJy beam$^{-1}$ for CS, SiO, and SO, respectively).
The positions of the SVS13-A, VLA3, SVS13-B, and SVS13-C continuum peaks are marked by black or white  stars (Table \ref{tab:continuum}). The synthesised beams are shown in the bottom left panel, and the dashed circles delimitate the FoV
(Tab. \ref{tab:lines}). The black triangles in  the top panels indicate the positions of the HH8, HH9, HH10, and HH11 Herbig-Haro objects \citep[e.g.][]{Bally1996}.
} \label{fig:cs-channels}
\end{center}
\end{figure*}

For setup 2, we used the Polyfix correlator, which covers the SiO(2--1)
frequency with the 80--88 GHz spectral band and has a resolution of 2 MHz ($\sim$6.8 km s$^{-1}$). 
The bandpass was calibrated using the same procedures and calibrators as used for setup 1, with a similar uncertainty on the absolute flux ($\leqslant$10$\%$). 
The continuum (line-free) image has a clean beam of 1$\farcs$6 $\times$ 1$\farcs$1 (PA = 41$\degr$), an rms noise of 100 $\mu$Jy beam$^{-1}$, and was published by \citet{Codella2021}.
The SiO image was produced by subtracting the continuum emission using natural weighting for a spatial resolution of 
1$\farcs$72 $\times$ 1$\farcs$20. We obtained an additional SiO(2--1)
image that was restored with a larger synthesised beam (2$\farcs$56 $\times$ 2$\farcs$14) for a proper comparison with the CH$_3$OH and SO data
(Table \ref{tab:lines}).
The rms noise in the line datacubes (without primary beam correction) at the SiO, CH$_3$OH, CS, and SO lines ranges from 0.6 mJy/beam to 2.0 mJy/beam, depending on the frequency and spectral resolution (see Table \ref{tab:lines}).

\begin{table*}
\caption{Molecular lines observed using NOEMA towards the SVS13 
protostellar system.}
\begin{tabular}{lcccccccc}
\hline
\multicolumn{1}{c}{Transition$^a$} &
\multicolumn{1}{c}{$\nu_{\rm 0}$ $^a$} &
\multicolumn{1}{c}{$E_{\rm up}$ $^a$} &
\multicolumn{1}{c}{$S\mu^2$ $^a$} &
\multicolumn{1}{c}{log(A$_{ij}$) $^a$} &
\multicolumn{1}{c}{d$V$} &
\multicolumn{1}{c}{rms} &
\multicolumn{1}{c} {FoV} &
\multicolumn{1}{c} {beam} \\
\multicolumn{1}{c}{ } &
\multicolumn{1}{c}{(GHz)} &
\multicolumn{1}{c}{(K)} &
\multicolumn{1}{c}{(D$^2$)} &
\multicolumn{1}{c}{ } &
\multicolumn{1}{c}{(km s$^{-1}$)} &
\multicolumn{1}{c}{(mJy beam$^{-1}$)} &
\multicolumn{1}{c}{($\arcsec$)} &
\multicolumn{1}{c}{($\arcsec$ $\times$ $\arcsec$, $\degr$)} \\
\hline
SiO(2--1) & 86.84696 & 6.3 & 19.2 & --4.5 & 6.90 & 1.2 & 57 & 1.72 $\times$ 1.20, 39$^b$ \\
 &  &  &  &  & 6.90 & 0.8 & 57 & 2.56 $\times$ 2.14, 31$^b$ \\
CH$_3$OH (2$_{\rm 1,2}$--1$_{\rm 1,1}$) A$^c$ & 95.91431 & 21 & 4.9 & --5.6  &  6.11 & 0.9 & 51 & 2.69 $\times$ 2.15, 31  \\
CH$_3$OH (2$_{\rm 1,2}$--1$_{\rm 1,1}$) E$^d$ & 96.73936 & 13 & 4.8 & --5.6  & 0.48  & 2.0  & 51 & 2.72 $\times$ 2.14, 22 \\ 
CH$_3$OH (2$_{\rm 0,2}$--1$_{\rm 0,1}$) A$^d$ & 96.74137 & 7 & 6.5 & --5.5  &  0.48 & 2.0  & 51 & 2.72 $\times$ 2.14, 22 \\
CH$_3$OH (2$_{\rm 0,2}$--1$_{\rm 0,1}$) E$^d$ & 96.74455 & 20 & 6.5 & --5.5  & 0.48 & 2.0 & 51 & 2.72 $\times$ 2.14, 22 \\
CH$_3$OH (2$_{\rm 1,1}$--1$_{\rm 1,0}$) E$^c$ & 96.75550 & 28 & 5.0 & --5.5  &  0.48 & 2.0 & 51 & 2.72 $\times$ 2.14, 22 \\
CH$_3$OH (2$_{\rm 1,1}$--1$_{\rm 1,0}$) A$^c$ & 97.58280 & 22 & 4.8 & --5.6  & 6.21 & 0.9  & 50 & 2.69 $\times$ 2.15, 31 \\
CS (2--1) & 97.98095 & 7 & 7.6 & --4.8 & 6.10 & 1.2 & 50 & 2.68 $\times$ 2.13, 31  \\
SO (2$_{\rm 3}$--1$_{\rm 2}$) & 99.29987 & 9 & 6.9 & --4.9 & 5.90 & 0.7 & 50 & 2.65 $\times$ 2.10, 31 \\
\hline
\label{tab:lines}
\end{tabular}
\tablefoot{
For each line datacube, we report the spectral resolution (d$V$), r.m.s. per channel, FoV, and synthesised beam. $^a$ The spectroscopic parameters (line frequency, $\nu_{\rm 0}$, upper level energy, $E_{\rm up}$, line strength, $S\mu^2$, and Einstein coefficients, log(A$_{ij}$)) are taken from \citet{Manson1977} (SiO), \cite{Xu1997} (CH$_3$OH), \citet{Gottlieb2003} (CS), \citet{Klaus1996}, and \citet{Bogey1997} (SO), retrieved from the Cologne Database for Molecular Spectroscopy (CDMS) database \citep{Muller2005}. $^b$ Two SiO(2--1) datacubes were produced: one cube at the spatial resolution resulting from the $uv$ coverage of our observations, and a second cube restored with a larger synthesised beam for a proper comparison with the CH$_3$OH and SO data. $^c$ No detection along the outflows. $^d$ The line is blended at the current spectral resolution (see Fig. \ref{fig:spectra-peaks}).} \\
\end{table*}

\section{Continuum emission} 
\label{sec:results-continuum}

\begin{figure}
\begin{center}
\includegraphics[angle=0,scale=0.2]{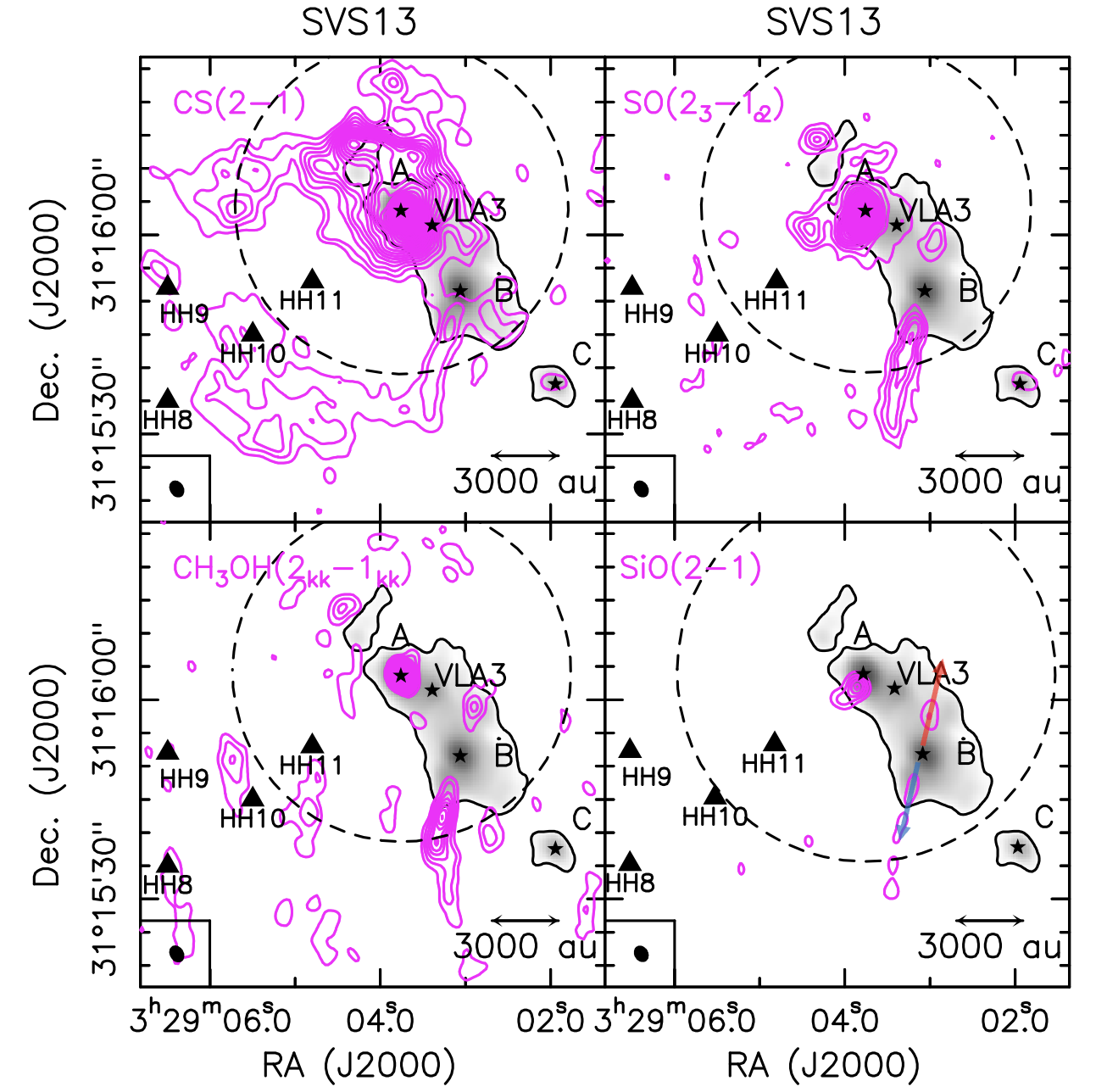}
\caption{SVS13 multiple system as traced by the integrated-intensity map (moment 0, magenta contours) of CS(2--1) (top left), SO(2$_3$--1$_2$) (top right),
CH$_3$OH(2$_{\rm kk}$--1$_{\rm kk}$) (Bottom-Left), and SiO(2--1) (Bottom-Right). The methanol emission is due to the lines of the 2$_{\rm kk}$--1$_{\rm kk}$ spectral pattern that is emitted at 96.7 GHz (see Tab. \ref{tab:lines}).
The maps were obtained by integrating on the velocity ranges [$-18.3$, $+6.1$] km s$^{-1}$ (CS), [$-106.3$, $+35.2$] km s$^{-1}$
(SO), [$-5.9$, $+19.8$] km s$^{-1}$ (CH$_3$OH), and [$-109.9$, $+48.9$] km s$^{-1}$ (SiO).
The first contours and steps are at 3$\sigma$ (1$\sigma$ = 16 mJy km s$^{-1}$ beam$^{-1}$ for CS, 20 mJy km s$^{-1}$ beam$^{-1}$ for SO, 
12 mJy km s$^{-1}$ beam$^{-1}$ for CH$_3$OH, and 
27 mJy km s$^{-1}$ beam$^{-1}$ for SiO).
We report the 3mm continuum image in greyscale (and black contour, 3$\sigma$) (Fig. \ref{fig:continuum}).
The continuum peaks of the SVS13-A, VLA3, SVS13-B, and SVS13-C protostars are 
marked by black stars (Table \ref{tab:continuum}).
The synthesised beams are shown in the bottom-left corners, and the dashed circles delimitate the FoV of the images 
(Tab. \ref{tab:lines}). The black triangles indicate the positions of the HH8, HH9, HH10, and HH11 Herbig-Haro objects \citep[e.g.][]{Bally1996}. The red and blue arrows in the bottom-right panel indicate 
the jet direction, as discussed in Sect. \ref{sec:jetB}.}
\label{fig:mom0}
\end{center}
\end{figure}

Figure \ref{fig:continuum} shows the SVS13 region as observed in dust continuum emission at 3 mm. It reveals the dust envelope that hosts the SVS13-A, SVS13-B, and VLA3  protostars, which are detected with a signal-to-noise ratio (S/N) of 60 at least. SVS13-A is the brightest at 3mm, and the binary components VLA4A and VLA4B are not separated at the current angular resolution. Although it is located outside the FoV (51$\arcsec$), SVS13-C is also clearly detected (S/N $\simeq$ 30). The J2000 coordinates of the  protostars and the peak intensities were obtained from a 2D fitting in the image plane, and they are reported in Tab \ref{tab:continuum}. 

Our continuum images can be compared with those obtained at 3mm, first using the IRAM-PdBI telescope \citep{Maury2019}, and then IRAM-NOEMA \citep{Codella2021,Bianchi2022a} at higher angular resolution 
($\geq$ 1$\farcs$8). 
When we take into account that our continuum maps are at a lower angular resolution, 
the peak flux intensities are also consistent.
No in-depth examination of the continuum data is presented here. Nonetheless, the protostellar positions were used to determine the sources that drive the outflows and jets, as detailed in the following sections.

\begin{table}
	\caption{Continuum peaks observed towards NGC1333-SVS13.}
	\begin{tabular}{lccc}
 \hline
Source  & $\alpha(\rm {J2000})$ & $\delta(\rm {J2000})$ & $I_{\rm 3mm}$ \\
& (h:m:s) & ($^{\degr}$ $^{\arcmin}$ $^{\arcsec}$) & (mJy beam$^{-1}$) \\
\hline
SVS13-A  & 03:29:03.758 & +31:16:03.63 & 36 \\
VLA3 & 03:29:03.394 & +31:16:01.43 &  5 \\
SVS13-B   & 03:29:03.060 & +31:15:51.55 &  27 \\
SVS13-C & 03:29:01.942 & +31:15:37.54 &  3 \\
\hline
\end{tabular}
\tablefoot{Position and intensity of the continuum peaks observed using NOEMA at 3 mm using configurations C and D (see Fig. \ref{fig:continuum}). The r.m.s. is 70 $\mu$Jy beam$^{-1}$.}
\label{tab:continuum}
\end{table}

\section{Line emission} 
\label{sec:results-line}

\subsection{SVS13-A: Molecular envelope and hot corino}
\label{sec:envelope}

We detected and imaged the SiO, SO, CS, and CH$_3$OH emission lines with $E_{\rm up}$ in the 7--28 K range (Table \ref{tab:lines}). Figure \ref{fig:mom0}
shows the spatial distribution of the emission from the four species, integrated over the whole emitting velocity range (moment 0 maps), overlapped on the continuum image (Fig. \ref{fig:continuum}).
The maps were obtained by integrating in different velocity ranges for each molecule: [$-18.3$, $+6.1$] km s$^{-1}$ (CS), [$-106.3$, $+35.2$] km s$^{-1}$
(SO), [$-5.9$, $+19.8$] km s$^{-1}$ (CH$_3$OH), and [$-109.9$, $+48.9$] km s$^{-1}$ (SiO). The methanol emission refers to
the sum of the (2$_{\rm 0,2}$--1$_{\rm 0,1}$) E, (2$_{\rm 0,2}$--1$_{\rm 0,1}$) A, and (2$_{\rm 1,1}$--1$_{\rm 1,0}$) E emission lines (see Sect. \ref{sec:jetB}).

\begin{figure}
\begin{center}
\includegraphics[angle=0,scale=0.4]{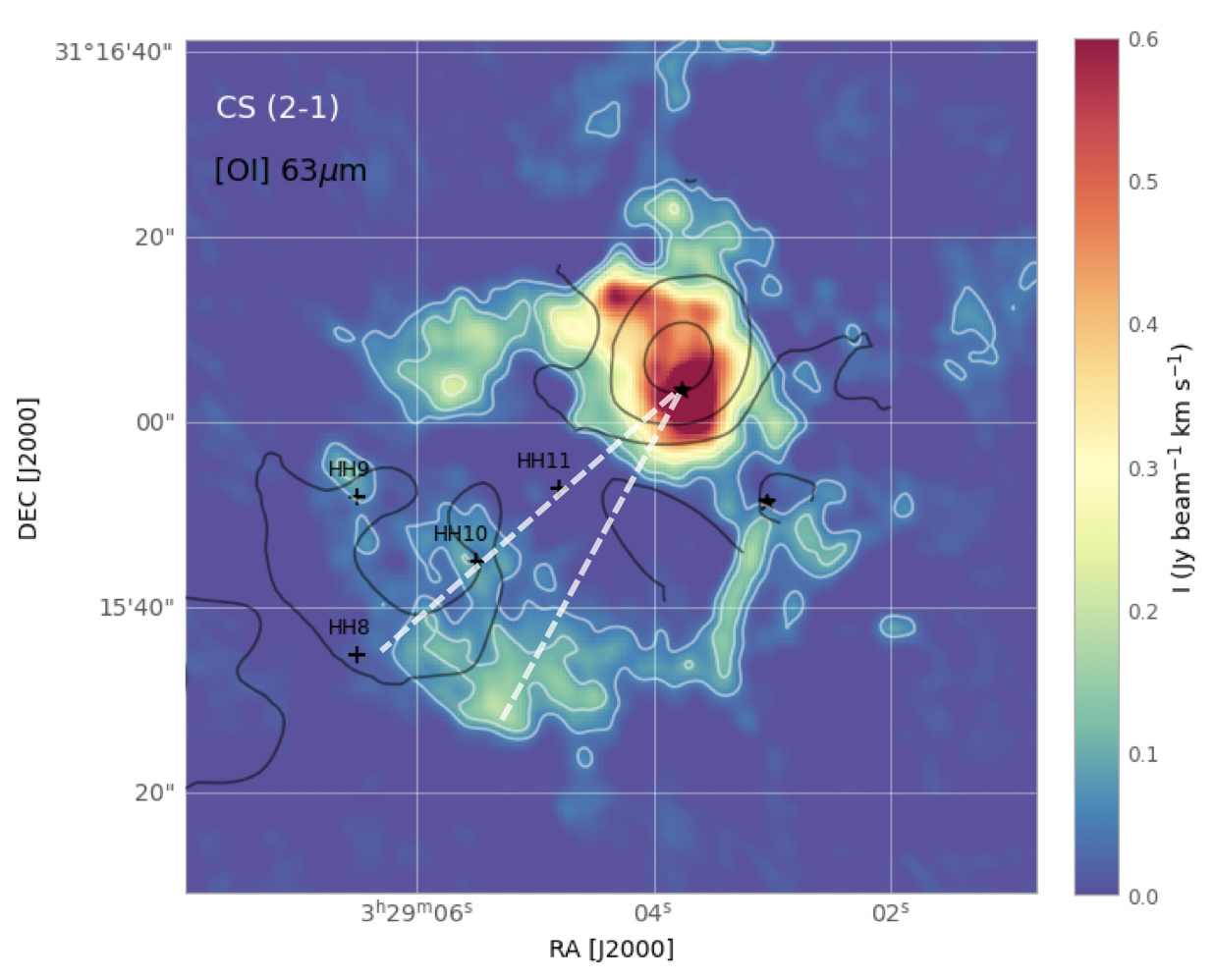}
\caption{SVS13 protostellar cluster: Comparison of our CS(2--1) NOEMA-SOLIS map (colour scale and white contours) and the spatial distribution of the [OI] gas observed at 63 $\mu$m with SOFIA (black contours; some contours are not closed due to
the FoV of the SOFIA observations). The contours correspond to three levels between (0.068-0.400) $\times$ 10$^{-13}$ erg s$^{-1}$ cm$^{-2}$, as in Figure 5 by \citet{Sperling2020}. The FWHM beam size of the SOFIA observations is 5$\farcs$4. The dashed lines indicate the direction of the two main jets: (i) PA $\simeq$ 130$\degr$ (connecting SVS13-A and HH11) and (ii) PA $\sim$ 155$\degr$ (traced by H$_2$, and CO bullets; see text).}\label{fig:sofia}
\end{center}
\end{figure}

The molecular envelope that hosts the SVS13-A
and VLA3 young stellar objects is  traced by CS(2--1) (Figure \ref{fig:mom0}-Upper Left), which overlaps the dust continuum emission in the region surrounding these sources (Fig. \ref{fig:continuum}).
The CS species was already employed to trace the SVS13 envelope 
using IRAM-30m observations at a spatial resolution of 10$\arcsec$--30$\arcsec$ \citep{Langer1996,Lefloch1998a}, as well as IRAM-NOEMA data at 1$\farcs$6 $\times$ 1$\farcs$1 resolution \citep{Codella2021}. While the single-dish surveys uncovered a very extended molecular structure that reached up to 10000 au in size, the higher-resolution IRAM-NOEMA CS(2--1) map by \citet{Codella2021} showed emission that was confined in a region of 450 au around SVS13-A. 
Our NOEMA map of CS(2--1) reveals a portion of the envelope with a size of about 5000 au, that is, intermediate between what was observed using single dishes ($\sim 10000$ au size) and the compact region around the SVS13A protostar highlighted by \citet{Codella2021} ($\sim 450$ au). Our map reveals the dense core, similarly to what was imaged with interferometers in other star-forming regions \citep[see, e.g., ][and references therein]{Ohashi2022}.
When we assume local thermodynamic equilibrium (LTE), optically thin emission, and a temperature for the envelope in the 20-40 K range 
\citep{Lefloch1998a}, the column density of CS at the position of its emission peak is $N_{\rm CS}$ $\simeq$ 2--3 $\times$ 10$^{14}$ cm$^{-2}$.

The SO(2$_3$--1$_2$) moment 0 map (Figure \ref{fig:mom0}-Upper Right) shows a structure of $\sim$ 2000 au around SVS13-A. This is smaller than the structure imaged by CS. 
This is consistent with the SO map reported by \citet{Codella2021}
using a smaller beam, $\sim$ 1$\farcs$4. Emission from the molecular envelope is not excluded, and neither is emission from the hot corinos (i.e. the chemically enriched region heated by  the protostar, with a temperature $\geq$ 100 K and a typical size of 100 au) that were detected around the two SVS13-A binary components \citep{Diaz2022,Bianchi2022a}. 
However, the SO emission clearly extends in the south-east direction, which also indicates emission from the molecular jet/outflow(s) associated with the HH7--11 chain. The SiO and SO jets driven by SVS13-A are reported in Sect. \ref{sec:outflowA},
but they will not be discussed in this paper, given the SVS13-A jets have been already published and analysed using different datasets in previous papers \citep[e.g.][and references therein]{Lefevre2017}.
When we assume a temperature $\geq$ 20 K \citep{Lefloch1998a}, LTE, and optically thin emission, the column density at the peak of the SO spatial distribution is $N_{\rm SO}$ $\geq$ 10$^{15}$ cm$^{-2}$.

Figure \ref{fig:mom0} (Lower Left) shows the spatial distribution 
of the CH$_3$OH(2$_{\rm kk}$--1$_{\rm kk}$) emission lines that emit close to 96.7 GHz (Tab. \ref{tab:lines}). Methanol emits in a compact region associated
with SVS13-A, which is currently spatially unresolved at the ($\sim$ 700 au) angular resolution. More precisely, Fig. \ref{fig:methanol-hotcorino} shows the methanol spectrum derived at the continuum peak position at the coordinates of SVS13-A (Tab. \ref{tab:continuum}). The four spectral lines of CH$_3$OH(2$_{\rm kk}$--1$_{\rm kk}$), which are associated with an upper energy level ($E_{\rm up}$) that ranges from 7 K and 28 K, exhibit a Gaussian profile. They have a full width at half maximum (FWHM) of 4.5$\pm$0.5 km s$^{-1}$ and display nearly identical peak emissions. This indicates large optical depths. These peaks, measured on the brightness temperature ($T_{\rm B}$) scale, range from 2.75 K to 2.94 K, with a margin of error of 0.05 K. These finding agree with 
the occurrence of hot-corino emission associated with the SVS13-A
binary system. Following the $\sim$ 0$\farcs$2 resolution images by \citet{Bianchi2022b}, the
region that emits in methanol has a size of $\sim$ 0$\farcs$4. The correction of the CH$_3$OH spectrum for the filling factor implies that the line intensities have to be multiplied by a factor
of $\sim$ 37, resulting in line peak temperatures between 102 K and 109 K.
Therefore, the observations indicate that the methanol emission probes an area of high optical depth with a surface temperature of $\geq$ 100 K.  
\citet{Bianchi2022b} analysed methanol maps with $E_{\rm up} >60$ K using the large velocity gradient (LVG) approach. They reported temperatures of 170$\pm$50 K, volume densities
higher than 2 $\times$ 10$^7$ cm$^{-3}$, and a CH$_3$OH column density $N_{\rm CH_3OH}$ = 5$\pm$1 $\times$ 10$^{18}$ cm$^{-2}$.
In the current datatset, with an assumed temperature of 100 K,
$N_{\rm CH_3OH}$ $\geq$ 6 $\times$ 10$^{17}$ cm$^{-2}$, which is consistent with prior observations.

\subsection{SVS13-A: Molecular shell}
\label{sec:shells}

The mass ejection from the SVS13-A binary system was
analysed by \citet{Lefevre2017} using IRAM-PdBI observations of the CO, SiO, and SO lines taken in the context of the CALYPSO survey, which they complemented by previous observations. The authors showed in their Fig. 5 that there are at least two jets: (i) 
H$_2$ \citep{Hodapp2014}, and CO high-velocity bullets reveal a wiggling jet with a mean PA of $\sim 155\degr$, and (ii) additional CO emission points in the HH7-11 direction with PA $=130\degr-140\degr$. 
As reported in Sect. \ref{sec:svs13}, the HH knots are not
spatially associated with the SiO that is produced by jet-driven shocks. 
Our channel maps of SiO and SO towards SVS13-A confirm the results by \citet{Lefevre2017}, hence high-velocity jet emission along PA$\sim 140\degr-155\degr$, and they are reported in Sect. \ref{sec:outflowA}. In this section, we focus on the detection of the shell-like structure in the SE direction observed in the CS(2--1) maps shown in Figs. \ref{fig:mom0}
and \ref{fig:cs-channels}.
Figure \ref{fig:sofia} shows the comparison between our CS(2--1) NOEMA-SOLIS map (colour scale) and the spatial distibution of the atomic oxygen gas observed at 63 $\mu$m with SOFIA (black contours) by \citet{Sperling2020}.
The CS and [OI] emissions are spatially displaced.
The emission from [OI] is extended, and it might trace the atomic component of the shocks associated with the HH7-11 chain.
Conversely, CS is notably shifted in the direction of the jet,  with a PA of 155$\degr$. In this direction, \citet{Hodapp2014} detected via H$_2$ observations a collection of expanding bubble segments that are indicative of low-velocity shocks. The prevailing theory suggests that they are the result of periodic explosive events that created multiple expanding bubbles throughout the outflow. Consequently, the detected CS structure resembling a shell is likely intertwined with material drawn from the surrounding molecular cloud.
The spectral resolution provided by the SOLIS observations does not allow us to perform a kinematic analysis to verify the expansion of the gas. Nonetheless, Fig. \ref{fig:cs-channels} indicates that the CS shell is blue-shifted and reaches velocities of up to $\sim -6$ km s$^{-1}$ with respect to the systemic velocity. This agrees with the direction of the blue-shifted southern jets that are driven by the SVS13-A binary system.

\subsection{SVS13-B: Protostellar jet}
\label{sec:jetB}

Our dataset enables us to map the protostellar jet driven by SVS13-B. Figure \ref{fig:mom0} shows emission by the bipolar jet in the NW-SE direction with a PA$\sim 167\degr$, which agrees with what was estimated based on previous maps of SiO (2--1), and SiO (5--4) emission \citep{Bachiller1998,Podio2021}.  
For the first time, we report jet emission also in the CS, SO, and CH$_3$OH lines. 
The channel maps of CS (Fig. \ref{fig:cs-channels}) clearly delineate the SE blue-shifted lobe, which emits down to --20 km s$^{-1}$ relative to the systemic velocity of +8.5 km s$^{-1}$.

The channel maps of the CS, SiO, SO, and CH$_3$OH emission are 
shown in Figs. \ref{fig:cs-channels}, and \ref{fig:ch3oh-channels}. They reveal the red- and blue-shifted jet lobes. 

\begin{figure}
\begin{center}
\includegraphics[angle=0,scale=0.6]{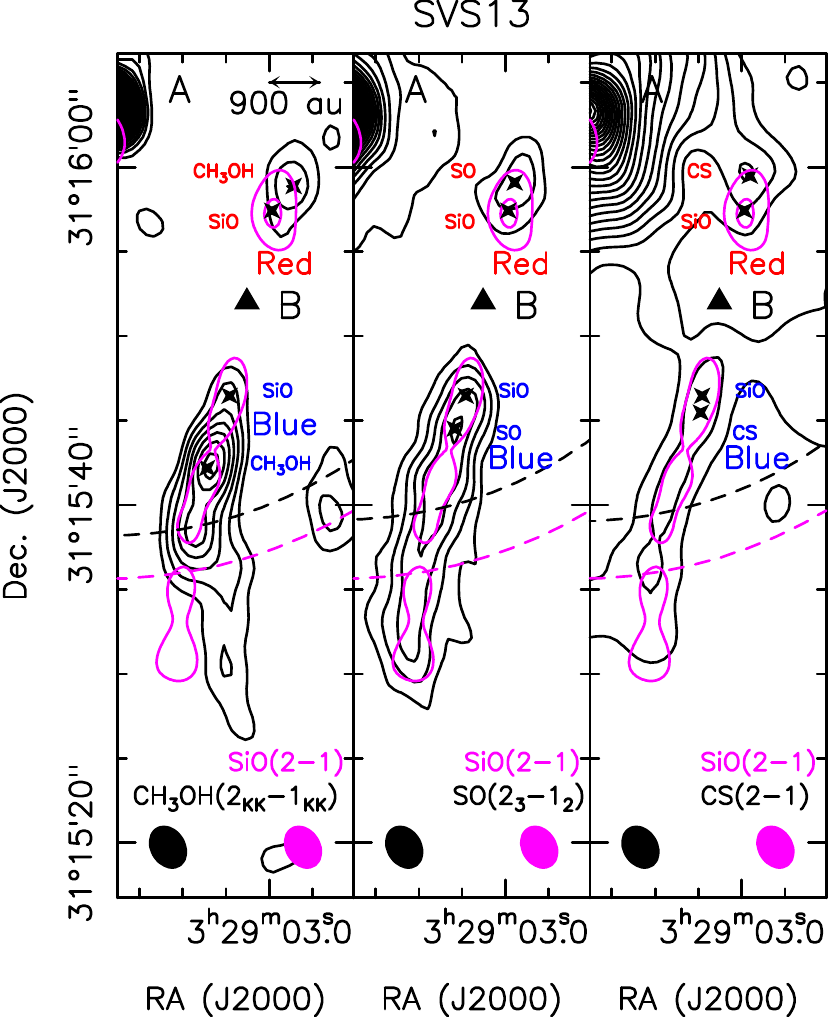}
\caption{Red- and blue-shifted bipolar SVS13-B jet as traced by SiO(2--1) is shown in magenta and is compared with CH$_3$OH(2$_{\rm kk}$--1$_{\rm kk}$) (Left panel), SO(2$_3$--1$_2$) (Middle panel), and CS(2--1) (Right panel).
The methanol emission is due to the lines of the 2$_{\rm kk}$--1$_{\rm kk}$ spectral pattern that emits at 96.7 GHz (see Tab. \ref{tab:lines}).
The SiO and SO maps were obtained by integrating on the following velocity ranges with respect to the systemic velocity \citep[+8.5 km s$^{-1}$;][]{Podio2021}: $\pm$ 33.8 km s$^{-1}$ (SiO), $\pm$ 23.6 km s$^{-1}$ (SO), and from
--18.3 km s$^{-1}$ to +6.1 km s$^{-1}$ (CS) (see Fig. \ref{fig:cs-channels}).
The first contours and steps are at 3$\sigma$, which corresponds to 
72 mJy km s$^{-1}$ beam$^{-1}$ (SiO), 45 mJy km s$^{-1}$ beam$^{-1}$ (CH$_3$OH), and 63 mJy km s$^{-1}$ beam$^{-1}$ (SO, and CS). 
The positions of the B continuum peak are
marked by a black triangle (Table \ref{tab:continuum}).
The crosses mark for each species 
the peaks of the red- and blue-shifted emission
(see labels).
The synthesised beams  are shown in the bottom-left corners, and the dashed circles delimitate the FoV (Tab. \ref{tab:lines}). For SiO, the dataset with the beam of 2$\farcs$56 $\times$ 2$\farcs$14 was used for a proper comparison with SO, CS, and CH$_3$OH.} \label{fig:comparison}
\end{center}
\end{figure}

\begin{figure*}
\begin{center}
\includegraphics[angle=0,scale=0.6]{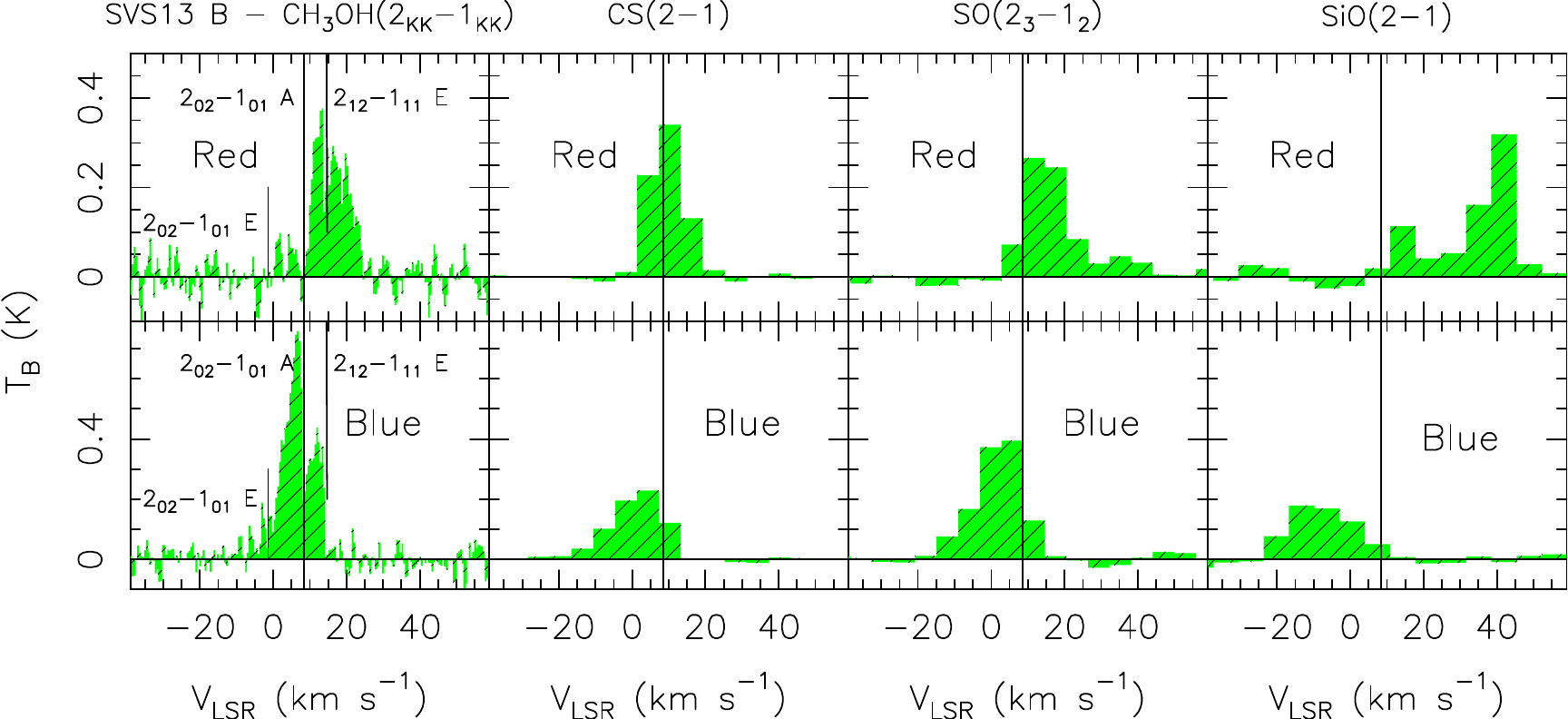}
\caption{From left to right: Spectra in brightness temperature ($T_{\rm B}$ scale) of CH$_3$OH(2$_{\rm kk}$--1$_{\rm kk}$), CS(2--1), SO(2$_3$--1$_2$),
and SiO(2--1) extracted at the positions of the emission peaks along the red- (top panels) and blue-shifted (bottom panels) lobes 
(see the red and blue crosses in Fig. \ref{fig:comparison}).
For SiO, the dataset with the beam of 2$\farcs$56 $\times$ 2$\farcs$14  was used (see Tab. \ref{tab:lines}).
The transitions producing the emission methanol profiles are labelled, and the corresponding frequencies are marked by vertical black segments (see Table \ref{tab:lines}). The CH$_3$OH spectra are centred at the frequency of the 2$_{\rm 0,2}$--1$_{\rm 0,1}$ A transition: 96741.38 MHz.
The vertical black lines at $+8.5$ km s$^{-1}$ indicate the systemic velocity \citep{Podio2021}.} \label{fig:spectra-peaks}
\end{center}
\end{figure*}

\subsection{Spatio-kinematical properties of the jet} 
\label{sec:spatio-kin}

The moment 0 maps in Fig. \ref{fig:comparison} show that SiO, CS, SO, and CH$_3$OH probe different regions of the SVS13-B protostellar jet. The molecular species peak at different positions\footnote{All the emission peaks have an S/N $\geq$ 6. The only exception is represented by the SiO red-shifted peak, revealed with an S/N = 5.}  along the jet (reported in Tab. \ref{tab:offset}), that is, the SiO peaks are located closer to the protostar, at a distance of $\sim$ 1600 au, while SO, CS, and CH$_3$OH peak at larger distances, $\sim$ 2000--2850 au. 
The implication of this spatial segregation is discussed in Sect. \ref{sec:discussion}. 

\begin{table*}
	\caption{Positions of the SiO, SO, CS, and CH$_3$OH emission peaks along the SVS13-B jet.}
	\begin{tabular}{lccccc}
 \hline
Peaks & $\Delta$$\alpha$, $\Delta$$\delta$ & $d$ & $V_{\rm peak}$--$V_{\rm sys}$ & $\int T_{\rm B} dV$ & $N$$^{a}$ \\
 & ($^{\arcsec}$,$^{\arcsec}$) & (au) & (km s$^{-1}$) & (K km s$^{-1}$) & (cm$^{-2}$) \\
\hline
SiO--red  & --0$\farcs$57,+5$\farcs$88 & 1773 & 
+33.4(3.5) & 5.1(0.2) & 5.2(0.2) $\times$ 10$^{13}$ \\
SiO--blue & +0$\farcs$96,--5$\farcs$08 & 1546 &
--21.8(3.5) & 4.0(0.2) & 4.2(0.2) $\times$ 10$^{13}$ \\
SO--red   & --0$\farcs$80,+7$\farcs$55 & 2270 & +3.5(3.0) & 4.6(0.2) & 3.8(0.2) $\times$ 10$^{14}$ \\
SO--blue & +1$\farcs$37,--7$\farcs$03 & 2142 & --2.8(3.0) & 6.7(0.2) & 5.5(0.2) $\times$ 10$^{14}$ \\
CS--red   & --1$\farcs$00,+7$\farcs$95 & 2401 & +1.7(3.0) &  4.3(0.2) &  8.6(0.4) $\times$ 10$^{13}$ \\
CS--blue & +0$\farcs$09,--6$\farcs$08 & 1818 & --4.1(3.0) &  4.2(0.2) &  8.4(0.4) $\times$ 10$^{13}$ \\
CH$_3$OH--red   & --1$\farcs$35,+7$\farcs$34 & 2231 & +4.0(0.3)$^b$ & 3.4(0.2)$^b$ & 1.6(0.2) $\times$ 10$^{15}$ \\
CH$_3$OH--blue & +1$\farcs$77,--9$\farcs$36 & 2848 & --1.7(0.3)$^b$ & 6.4(0.2)$^b$ & 4.7(0.2) $\times$ 10$^{15}$ \\
\hline
\label{tab:offset}
\end{tabular}
\tablefoot{Emission peaks of SiO(2--1), SO(2$_3$--1$_2$), CS(2--1), and CH$_3$OH(2$_{\rm kk}$--1$_{\rm kk}$). We report the spatial offsets ($\Delta$$\alpha$, $\Delta$$\delta$) and distance ($d$) with respect to the coordinates of the driving protostar (see Tab. \ref{tab:continuum}), the peak
 velocity in the spectra extracted at the emission peaks ($V_{\rm peak}$--$V_{\rm sys}$), the velocity integrated intensities ($\int T_{\rm B} dV$), and the column densities ($N$). 
$^{a}$ Assuming a kinetic temperature of 100 K.
$^b$ The methanol velocity integrated emission refers to
the sum of the blended (2$_{\rm 0,2}$--1$_{\rm 0,1}$) E, (2$_{\rm 0,2}$--1$_{\rm 0,1}$) A, and (2$_{\rm 1,1}$--1$_{\rm 1,0}$) E emission lines (see Sect. \ref{sec:jetB}). The peak velocity
was measured from the (2$_{\rm 0,2}$--1$_{\rm 0,1}$) A profile.} \\
\end{table*}

The different spatial distribution of SiO with respect to SO, CS, and CH$_3$OH is accompanied by a different velocity distribution.
Figure \ref{fig:cs-channels} shows the channel maps of the SiO(2--1) red- and blue-shifted emission overlaid on the the 3mm continuum image. The emission close to SVS13-A is well known \citep[see e.g.][]{Lefevre2017,Podio2021},and is described in Sect. \ref{sec:outflowA}. 
In addition to the jet from SVS13-A, a clumpy jet is observed south of the protostellar source SVS13-B. It reaches velocities that are blue-shifted by approximately 30 km s$^{-1}$ (projected velocity). The red-shifted counterpart (see Fig. \ref{fig:cs-channels}) consists of only one clump that emits at a velocity of 30--35 km s$^{-1}$. Our maps complete what missed by \citet{Bachiller1998}, who, due to a limited spectral bandwidth, detected
SiO(2-1) emission at velocities up $\pm$ 17 km s$^{-1}$ from the systemic velocity.
Within the central 500 astronomical units surrounding the protostar, no emission was observed. However, SiO(5--4) emission was identified by \citet{Podio2021} in the vicinity of SVS13-B. This indicates potential excitation effects, which we explored in Sect. \ref{sec:discussion}. 

The red- and blue-shifted jet is also well outlined by the 
SO(2$_3$--1$_2$) channel maps (Fig. \ref{fig:cs-channels}). The emission is bright ($\sim$ 20$\sigma$) and has lower radial velocities than exhibited by SiO, with (projected) terminal velocities of $\pm$ 20 km s$^{-1}$ with respect to the cloud velocity. In contrast to SiO, which peaks at a high velocity ($\sim 20-30$ km\,s$^{-1}$), the SO emission peaks at velocities of $5-10$ km\,s$^{-1}$ with respect to $V_{\rm sys}$ (see Fig. \ref{fig:spectra-peaks}). The jet was previously imaged in SO by \citet{Podio2021} using the 6$_5$--5$_4$ line at 1.3mm, but the authors reported weak ($\geq$ 5$\sigma$) and slow (up to $\pm$ 5 km s$^{-1}$) emission that did not distinctly outline the jet structure. 

The CS(2--1) channel maps (Fig. \ref{fig:cs-channels}) also reveal a molecular jet. In this instance, the red-shifted knot is visible, but it is located within the protostellar envelope. As a result, the emissions from the two components are likely mixed. 
Conversely, the blue-shifted lobe is clearly outlined 
up to a 6$\sigma$ dynamical range, and (projected) terminal velocities similar to those of the SO emission, that is, 
$\pm$ 20 km s$^{-1}$ with respect to the cloud velocity.  

Finally, Fig. \ref{fig:ch3oh-channels} shows the kinematics of the
protostellar jet B as traced by CH$_3$OH. 
In the case of methanol, it is challenging to investigate the jet kinematics because the different components of the CH$_3$OH emission are blended. To minimize this effect, Fig. \ref{fig:ch3oh-channels} shows the red-shifted emission of the 2$_{\rm 1,2}$--1$_{\rm 1,1}$ E (Fig. \ref{fig:ch3oh-channels}-Upper panels) and the blue-shifted emission of the 2$_{\rm 1,2}$--1$_{\rm 1,1}$ A (Lower panels).
The figure shows that methanol emits at projected velocities lower than those showed by SO and SiO. More specifically, the red-shifted component consists of an NW compact clump that emits up to $\simeq +5$ km s$^{-1}$ with respect to $V_{\rm sys}$. The blue-shifted counterpart shows a more complex structure. In addition to the SE lobe that is also detected in SiO and SO, another feature curves southward. This feature is situated more than 15 arcseconds away from SVS13-B, and its radial velocities are lower than $5$ km s$^{-1}$. The comparison of the SiO, CS, SO, and CH$_3$OH emissions is discussed in Sect. \ref{sec:discussion}.

\section{Discussion} \label{sec:discussion}

\subsection{Chemical stratification in the SVS13-B jet} \label{sec:differentregions}

By combining the spatial distribution of the emission lines that is shown by the moment 0 maps in Fig. \ref{fig:comparison}, with their velocity distribution that is shown in the spectra in Fig. \ref{fig:spectra-peaks}, we discuss the regions of the SVS13-B jet that are traced by each  molecular species.
In this context, Tab. \ref{tab:offset} lists the spatial offsets 
of the red- and blue-shifted emission peaks with respect to the coordinates of the driving protostar and the corresponding velocity peaks.

The observed species probe different gas components, which are at different
velocities. The SiO(2-1) spectra extracted at the emission peaks along the SVS13-B jet (see Fig. \ref{fig:spectra-peaks}) cover a wide range of radial velocities (up to $\sim 40-45$ km\,s$^{-1}$ with respect to systemic), as expected in the case of a molecular jet with typical deprojected velocity of 100 km\,s$^{-1}$ \citep[e.g.][]{Podio2021}. 
Assuming that the jets propagate perpendicular to the disk, and deprojecting for the disk inclination of $i \sim 71\degr$, as measured by \citet{Segura-Cox2016}, we estimate that the SiO jet propagates with a velocity of $\sim92$ km\,s$^{-1}$, which agrees with the assumed velocity of 100 km\,s$^{-1}$ reported by \citet{Podio2021}.

The SO(2$_3$--1$_2$) line also emits at high velocities and reaches $-20$ 
km\,s$^{-1}$ and $+30$ km\,s$^{-1}$. Interestingly, the SO emission at the red-shifted peak shows faint emission up to $+40$ km\,s$^{-1}$, but with 
an S/N close to 3$\sigma$. However, the bulk of SO emission is at lower velocities than SiO. It peaks at  velocities of $\sim \pm3$ km\,s$^{-1}$ with respect to the systemic velocity, which is an order of magnitude lower than the SiO velocities.
The CS(2--1) profile extracted at the blue-shifted peak is very similar
to that of SO, and the red-shifted emission, as reported in Sect. \ref{sec:spatio-kin}, might be contaminated by emission at systemic velocity from the extended molecular envelope, as illustrated in Fig. \ref{fig:cs-channels}.
Finally, the CH$_3$OH(2$_{\rm kk}$--1$_{\rm kk}$) emission along the SVS13-B jet axis is associated with low velocities that peak at 2--4 km\,s$^{-1}$ with respect to the systemic velocity of +8.5 km\,s$^{-1}$.

In addition to the difference in velocities, as reported in Sect. \ref{sec:jetB}, SiO, SO, CS, and CH$_3$OH peak at different positions along the SVS13-B protostellar jet. As reported in Tab. \ref{tab:offset}, SiO peaks at a distance of $\sim$ 1600 au from the protostars, SO and CS at $\sim$ 2200 au, and CH$_3$OH shows the peak emission at $\sim2200-2800$ au.
Moreover, different species show different transversal sizes, that is, they probe different transversal layers of the jet.
Methanol emission comes from
a wider-angle jet layer ($\simeq$ 750 au wide) than  SiO, which is more collimated, with a width smaller than 450 au (taking into account the two SiO datasets reported in Tab. \ref{tab:lines}). 

Silicon monoxide is widely accepted as the best 
tracer of protostellar jets 
\citep[see e.g.][and references therein]{Frank2014,Podio2021}.
Silicon monoxide is enhanced in shocks due to the release of Si, caused by the sputtering of the grain refractory cores, which rapidly reacts with O$_2$ or OH to forms SiO; and/or due to the direct release of SiO caused by the 
vaporisation of the grain mantles \citep{Guillet2011, lesaffre_low_2013, nguyen-luong_low_2013,de_simone_tracking_2022}. Shock velocities higher than $\sim$ 25 km s$^{-1}$ are necessary to release Si or SiO from the grain cores/mantles \citep[e.g.,][]{caselli_grain-grain_1997, schilke_sio_1997, gusdorf_sio_2008a, gusdorf_sio_2008b, guillet_shocks_2011}. 

Different from SiO, methanol is thought to be exclusively formed on grain surfaces \citep[e.g.,][]{watanabe_efficient_2002, rimola_combined_2014}. The release in the gas phase may occur in different environments. Firstly, CH$_3$OH can be released in the gas phase when the dust is heated to temperatures of $\simeq$ 100 K, which causes the submination of the icy grain mantles. 
This is the case reported in Sect. \ref{sec:envelope},
where methanol probes the hot corino around SVS13-A. 
Alternatively, CH$_3$OH can be released in gas in shocked regions, 
for example, in the shocks along protostellar jets.
Relatively low-velocity shocks, $\leq$ 10 km s$^{-1}$, are adequate to release species that were previously frozen onto the grain mantles
\citep{jimenez-serra_parametrization_2008, Guillet2011, 
lesaffre_low_2013,
nguyen-luong_low_2013,de_simone_tracking_2022}. 

The origin of SO and CS is linked to the chemistry of S-bearing species. Despite many dedicated studies \citep[see e.g. the GEMS\footnote{https://www.oan.es/gems/} project,][]{Fuente2023}, this is still far from fully understood.
More specifically, grain surface chemistry and gas-phase reactions (when conditions allow for grain products to be released into the gas) depend on where sulfur is bound on the dust.
The advent of the James Webb Space Telescope (JWST),
which observes in the mid-infrared, will
certainly provide a deeper knowledge of the dust composition in star-forming regions \citep[see e.g. the JOYS+ project,][]{Gelder2024}.
In shocked regions, the abundance of gaseous S-bearing species increases drastically 
\citep[e.g.,][]{Charnley1997,Bachiller2001,Imai2016,Taquet2020,Feng2020}.
So far, the main candidates to be the principal reservoir of sulfur on dust mantles are H$_2$S and OCS \citep[see e.g., ][and references therein]{Podio2014,Boogert2015,Taquet2020,Codella2021,Yang2021}.
Very recently, \citet{Rocha2024} detected SO$_2$ using JWST observations of protostellar regions. After the mantle products are released in the gas phase, the S-species boost a chemistry that leads to a significant increase in SO and SO$_2$. SO was even used 
as protostellar jet tracer in combination with SiO in the statistical study by \citet[][and references therein]{Podio2021}.
CS is also expected to increase the abundance in molecular outflows
in which the material is shocked. The pioneering project by \citet{Bachiller1997}, who mapped the
archetypical L1157 outflow with the IRAM 30m telescope, showed that CS in shocks can increase its abundance by two orders of magnitude with respect to the protostellar envelope. However, to our knowledge, no dedicated interferometric studies of CS in protostellar jets have been performed.

To summarise, current multi-species observations reveal a stratified structure in the SVS13-B jet: (i) a collimated ($\le 450$ au) high-velocity component (deprojected velocity around 100 km s$^{-1}$) that traces the region within approximately 1800 au from the driving star, (ii) a slower component (around 10 km s$^{-1}$) that is wider-angle ($\sim 750$ au) and extend to larger distances, traced by methanol (see Fig. \ref{fig:comparison}), and (iii) an intermediate component detected using SO and CS, which moves at a deprojected velocity of up to 50 km s$^{-1}$. 
These findings may indicate that methanol in shocked gas is the first element to return in the gas phase (i.e. it is abundant on the dust surface), probably because the catastrophic CO freeze-out required for efficient CH$_3$OH formation only occurs in the late stages of dense cores just before star formation.

High spatial resolution imaging (down to 10 au) of the SVS13-B jet using various molecules, including those analysed in this paper, will help us to reconstruct the complex structure of shocked material in protostellar jets.

\subsection{Column densities and excitation along the SVS13-B jet} \label{sec:excitation}

When we assume LTE conditions, optically thin emission, and a temperature of 100 K (as assumed by \citealt{Podio2021}), the SiO, SO, and CS column densities at the red- and blue-shifted peaks (see Tab. \ref{tab:offset}) are $N_{\rm SiO}$ $\simeq$ 4--5 $\times$ 10$^{13}$ cm$^{-2}$, $N_{\rm SO}$ $\simeq$ 4--6 $\times$ 10$^{14}$ cm$^{-2}$, and $N_{\rm CS}$ $\simeq$ 8 $\times$ 10$^{13}$ cm$^{-3}$.

When they are observed with a high-velocity resolution (0.48 km s$^{-1}$), the transitions producing the CH$_3$OH(2$_{\rm kk}$--1$_{\rm kk}$) emission pattern are spectrally blended. 
The 2$_{\rm 0,2}$--1$_{\rm 0,1}$ A and 2$_{\rm 1,2}$--1$_{\rm 1,1}$ E lines are definitely brighter than the 2$_{\rm 0,2}$--1$_{\rm 0,1}$ E line, which is associated with a critical density that is lower by about one order of magnitude
\citep{Lin2022}.
In Fig. \ref{fig:spectra-peaks} the CH$_3$OH spectra are centred at the frequency of the 2$_{\rm 0,2}$--1$_{\rm 0,1}$ A transition (96741.38 MHz). The other lines of the pattern (see Table \ref{tab:lines}) are marked by vertical black segments. The complexity of the red-shifted profile precludes the possibility of a realible fitting. Conversely,
within the CH$_3$OH blue profile, it is possible to distinguish the 2$_{\rm 0,2}$--1$_{\rm 0,1}$ A and 2$_{\rm 1,2}$--1$_{\rm 1,1}$ E peaks. The lines are blue-shifted and exhibit a blue wing. The overall spectral pattern was fit by assuming two Gaussian components with a FWHM of 3 km s$^{-1}$ at different velocities for each line: one component shifted by --2.0 km s$^{-1}$, and the other component shifted  by --5.3 km s$^{-1}$.
Figure \ref{fig:methanol-spectralfit} shows the four components we used to obtain the overall fit. From the brightness of the two transitions, we obtained an approximate estimate of the excitation temperature (20$\pm$10 K) and of the CH$_3$OH column density $N_{\rm CH_3OH}$ $\simeq$ 3 $\times$ 10$^{14}$ cm$^{-2}$. again adopting LTE, an CH$_3$OH-A/CH$_3$OH-E abundance ratio equal to 1, and optically thin emission. 

In order to constrain the physical conditions associated with the SiO emission in the SVS13-B jet, we compared the J = 2--1 line intensity with that of SiO(5--4) that was reported in the context of the CALYPSO IRAM PdBI Large Program by \citet{Podio2021}. For a proper comparison, we smoothed 
the SiO(5--4) image by \citet{Podio2021}, which was originally obtained with
a beam of 0$\farcs$67 $\times$ 0$\farcs$50, to the spatial resolution of our SiO(2--1) map (1$\farcs$72 $\times$ 1$\farcs$20, 39$\degr$). 
Figure \ref{fig:calypso-composed} (Left panel) 
compares the integrated-velocity maps of
the SiO(2--1) and (5--4) maps. 
The maps are consistent, which allows us to compare their morphologies inside the FoV of the SiO(5--4) image ($\sim$ 20$\arcsec$), which is smaller than that of SiO(2--1) (57$\arcsec$). 
Notably, the higher excitation transitions $J$ = 5--4 ($E_{\rm up}$ = 31 K compared to $J$ = 2--1, 6 K) effectively trace the inner 1000 au of the SVS13-B jet, while SiO(2--1) is revealed only at approximately 1600 au from the driving protostar. This suggests higher excitation conditions of the SiO gas close to the protostar. A qualitative estimate of the excitation conditions can be made using the SiO(2--1) and SiO(5--4) profiles extracted at the same position (see Fig. \ref{fig:calypso-composed}, Right panels). While in the red-shifted peak the 
SiO(2--1)/SiO(5--4) intensity ($T_{\rm B}$) ratio appears to be almost constant, the blue-shifted profiles show that the higher the velocity, the higher the excitation. In other words, the excitation of the SiO gas appears to increase with the emitting velocity and to decrease with the distance from SVS13-B. 

When we adopt the collisional coefficients by \citet{Balanca2018}, the critical densties at 100 K for the $J$ = 2--1 and 5--4 transitions are
$\sim$ 2 $\times$ 10$^{5}$ cm$^{-3}$ and $\sim$ 2 $\times$ 10$^{6}$ cm$^{-3}$, respectively. The difference in critical densities led us to discuss 
the line ratio and the SiO(5--4) intrinsic brightness
adopting a LVG approach, as was reported by \citep{Cabrit2007} for
the investigation of the HH 212 SiO jet, for example (see their Fig. 4, Upper panel).
When we assume a kinetic temperature of 100 K, the combination of SiO(5--4)/SiO(2--1) $\leq$ 1 and $T_{\rm B}$(5--4) $\simeq$ 0.5-0.8 (found at the highest SiO velocities) implies
volume densities ($n_{\rm H_2}$) higher than 3 $\times$ 10$^{5}$ cm$^{-3}$,
as is indeed expected for SiO jets \citep[e.g., ][and references therein]{Cabrit2012,Podio2021}. Consequently, the SiO column densities 
should be higher than 10$^{13}$ cm$^{-2}$. This number is consistent with the estimates obtained using an LTE approach (see Tab. \ref{tab:offset}). These findings clearly need to be verified by 
multi-line SiO observations. 

\subsection{The abundance ratio [SO]/[CS] in the SVS13-A molecular shell}\label{sec:cs-so}

The NGC1333 region was extensively mapped in multi-line CS emission by \citet{Langer1996} and \citet{Lefloch1998a} using the IRAM 30-m telescope. Specifically, \citet{Lefloch1998a} analysed CS(5--4) and continuum emission at 1.3mm to identify a shell-like structure called Cav1, located south of SVS13-A and shaped by the outflow. The IRAM 30-m single-dish telescope provides a spatial resolution of 10$\arcsec$, but the current CS IRAM-NOEMA image has confirmed the presence of the CS shell.
Interestingly, while we observe both CS and SO in the SVS13-B jet (and in the SVS13-A molecular envelope), we detect CS in the shell, but SO is only tentatively detected at the 3$\sigma$ level. 
Figure \ref{fig:spectrashell} shows the CS(2--1) and SO(2$_3$--1$_2$) spectra extracted at the position 
in the SVS13-A southern molecular shell where the CS emission peaks:
$\alpha(\rm {J2000})$ = 03$^{h}$ 29$^{m}$ 05$\fs$317, $\delta(\rm {J2000})$ = +31$^{\degr}$ 15$^{\arcmin}$ 30$\farcs$09.
The excitation conditions of the gas in the shell are not known.
On the other hand, the upper-level excitation of the observed CS(2--1) and SO(2$_3$--1$_2$) transitions are similar ($E_{\rm up}$ = 7 K and 9 K for CS, and SO, respectively). In addition, the critical densities (at temperatures lower than 80 K) of the two transitions are also similar \citep[see e.g.][]{Shirley2015}: 
0.7--1 $\times$ 10$^{5}$ cm$^{-3}$ for CS(2--1), and 
$\sim$ 3 $\times$ 10$^{5}$ cm$^{-3}$ for SO(2$_3$--1$_2$).
In light of this, we determined the CS and SO column densities in the shell
at the position of the CS line peak, assuming conservatively LTE conditions, temperatures in the 20--50 K range, and optically thin emission: 
$N_{\rm CS}$ = 2--4 $\times$ 10$^{13}$ cm$^{-2}$, and $N_{\rm SO}$ = 1--2 $\times$ 10$^{13}$ cm$^{-2}$. Therefore, the abundance ratio, as derived from the column densities [SO]/[CS] = $N_{\rm SO}$/$N_{\rm CS}$, is $\simeq$ 0.5.

Although a word of caution is needed because the shell that is detected at low velocities is outside the FoV of the images, the CS line is much brighter than the SO line in the shell, and the [SO]/[CS] abundance ratio is lower by one order of magnitude than that derived in the jet ([SO]/[CS] $\simeq$ 6; see Tab. \ref{tab:offset} and Fig. \ref{fig:spectra-peaks}). 
This agrees with   \citet{Wakelam2005}, who estimated the [SO]/[CS] in the protostellar region NGC1333-IRAS2 and reported a ratio of 1-3 at velocities close to the systemic velocity and a higher value of [SO]/[CS]$\sim3-30$ at the high velocities of the IRAS2A jet. 
The reasons for this finding are not straightforward because the abundance ratio [SO]/[CS] depends on many parameters, including, for example, the density and temperature of the gas and the composition of the grain mantles \citep[e.g.][]{Wakelam2005}. 
In addition, some studies suggested that a higher  [SO]/[CS]  abundance ratio may be due to a local decrease in the C/O ratio that favours the formation of SO over CS \citep{Semenov2018,Nilsson2000,LeGal2021,Bergin2024}.
In the case of the jet of SVS13-B, the gas-phase chemistry, including the C/O ratio, might be altered locally by the shocks along the jet \citep[e.g., ][]{Bachiller1997}.
In this context, the low [SO]/[CS] abundance ratio in the molecular shell might arise because the gas is  quiescent or because the shell is associated with a young shock in which the initial [SO]/[CS] value has not yet been significantly altered.

Finally, \citet{LeGal2019} reported that [SO]/[CS] is higher in diffuse gas (i.e. in the Horsehead region) than in denser regions (i.e. in disks).
In our study of the NGC1333 SVS13 protostellar cluster, however, we find the opposite, as the [SO]/[CS] ratio is higher in the jet, which is likely denser than the shell. However, this discrepancy may be reconciled when we speculate that SO formed at earlier times than CS and was therefore stored in a deeper layer of the grain mantles. It can therefore only be released in shocks or in irradiated regions. 

In this context, the displacement between the spatial distribution of CS and that of [OI] shown in Fig. \ref{fig:sofia} might be a signature of a different post-shock chemistry.
Clearly, this is a speculation based on the analysis of emission outside the FoV of the observations, and it can only be verified by further multi-line observations of CS and SO centred on the molecular shell to infer its excitation conditions.

\section{Summary and conclusions} \label{sec:conclusions}

The IRAM NOEMA Large Program SOLIS has conducted a survey at 3mm of SiO, SO, CS, and CH$_3$OH emission towards the NGC1333 SVS13 protostellar cluster. The SVS13-A system and the SVS13-B protostar were both imaged. The key findings are listed below.

\begin{itemize}

\item 
While the paper primarily focused on the SVS13-B jet, it also presented several findings related to SVS13-A. The CS and SO lines trace the molecular envelope, which is approximately 5000 au in size, with temperatures of 20--40 K. The envelope hosts the hot corino around SVS13-A, as indicated by CH$_3$OH. For the first time, a blue-shifted molecular shell was imaged in CS(2--1) in the direction of the jet driven by the SVS13-A system (position angle of 155$\degr$, as traced by high-velocity SiO and SO clumps). \\

\item 
The protostellar jet driven by the SVS13-B protostar was imaged in SiO and also in the SO, CS, and CH$_3$OH lines for the first time. The jet is aligned in the SE-NW  direction with a PA of 167$\degr$. 
The molecules
peak at different positions along the jet: SiO(2--1) peaks at $\sim$ 1600 au from SVS13-B, and SO(2$_{\rm 3}$--1$_{\rm 2}$), CS(2--1), and CH$_3$OH(2$_{\rm k,k}$--1$_{\rm k,k}$) peak at larger distances of $\sim$ 2000--2850 au.\\

\item 
The different spatial distribution of the observed species in the SVS13-B jet is accompanied by different velocity distributions.
The SiO emission peaks at $+35$ km s$^{-1}$ and $-20$ km s$^{-1}$ with respect to the systemic velocity in the red- and blue-shifted lobes, respectively. SO and CS peak at $\pm$ 10 km s$^{-1}$, and CH$_3$OH peaks at low velocities,  $\sim \pm$ 4 km s$^{-1}$.
Moreover, the emission is stratified transversally to the jet width. A collimated component (transversal size $\leq$ 450 au) is traced by SiO, which forms efficiently due to high-velocity shocks (above 25 km s$^{-1}$) that cause dust sputtering and vaporisation. A slower and wider component (transversal size $\sim$ 750 au) is traced by methanol. It is released from dust mantles at shock velocities below 10 km s$^{-1}$. CS and SO trace an intermediate component between the components traced by SiO and CH$_3$OH.\\

\item 
The SiO(2--1) map of the SVS13-B jet was compared with the previously published SiO(5--4) data. 
The higher excitation transition $J$ = 5--4 ($E_{\rm up}$ = 31 K) traces the inner 1000 au of the SVS13-B jet, while the lower excitation $J$ = 2--1 ($E_{\rm up}$ = 6 K) is detected only at $\sim 1600$ au from the driving protostar. 
Moreover, the spectra extracted at the emission peak along the blue-shifted lobe show higher SiO(5--4)/SiO(2--1) intensity ratios, and hence,  higher excitation, at higher velocities. Overall, our observations suggest that the excitation of the SiO gas is stronger close to the driving source SVS13-B and at high velocities.

\end{itemize}

To conclude, this paper contributes to filling the gap in the chemical richness of protostellar jets, which was so far only investigated in bright shocked regions 
\citep[e.g.][]{Codella2017,Tec2019,Desimone2020solis,Podio2021}.
We reported for the first time a comparison 
of the spatial and velocity distribution of the four detected tracers (SiO, SO, CS, and CH$_3$OH) in the protostellar jet, and we showed that the gas excitation is stratified in velocity and in chemistry. Clearly, multi-line and multi-tracer studies of a statistical sample are necessary to characterise the chemo-physical composition of protostellar jets.




\begin{acknowledgements}

We thank the anonymous referee for their instructive comments and suggestions. We are very grateful to all the IRAM staff, whose dedication allowed us to carry out the SOLIS project. 
This project has received funding from the EC H2020 research and innovation programme for: (i) the project "Astro-Chemical Origins” (ACO, No 811312), and (ii) the European Research Council (ERC) project “The Dawn of Organic Chemistry” (DOC, No 741002). 
ClCo, and LP acknowledge the PRIN-MUR 2020  BEYOND-2p (Astrochemistry beyond the second period elements, Prot. 2020AFB3FX), the project ASI-Astrobiologia 2023 MIGLIORA (Modeling Chemical Complexity, F83C23000800005), the INAF-GO 2023 fundings PROTO-SKA (Exploiting ALMA data to study planet forming disks: preparing the advent of SKA, C13C23000770005), the INAF Mini-Grant 2022 “Chemical Origins” (PI: L. Podio),
and the National Recovery and Resilience Plan (NRRP), Mission 4, Component 2, Investment 1.1, Call for tender No. 104 published on 2.2.2022 by the Italian Ministry of University and Research (MUR), funded by the European Union – NextGenerationEU– Project Title 2022JC2Y93 Chemical Origins: linking the fossil composition of the Solar System with the chemistry of protoplanetary disks – CUP J53D23001600006 - Grant Assignment Decree No. 962 adopted on 30.06.2023 by the Italian Ministry of Ministry of University and Research (MUR). EB aknowledges contribution of the Next Generation EU funds within the National Recovery and Resilience Plan (PNRR), Mission 4 - Education and Research, Component 2 - From Research to Business (M4C2), Investment Line 3.1 - Strengthening and creation of Research Infrastructures, Project IR0000034 – “STILES - Strengthening the Italian Leadership in ELT and SKA”.
\end{acknowledgements}

\bibliographystyle{aa} 
\bibliography{SVS13B-SOLIS} 


\begin{appendix}

\section{The CS and SO line profiles in the SVS13-A molecular shell}
\label{sec:shell}

\begin{figure}
\begin{center}
\includegraphics[angle=0,scale=0.4]{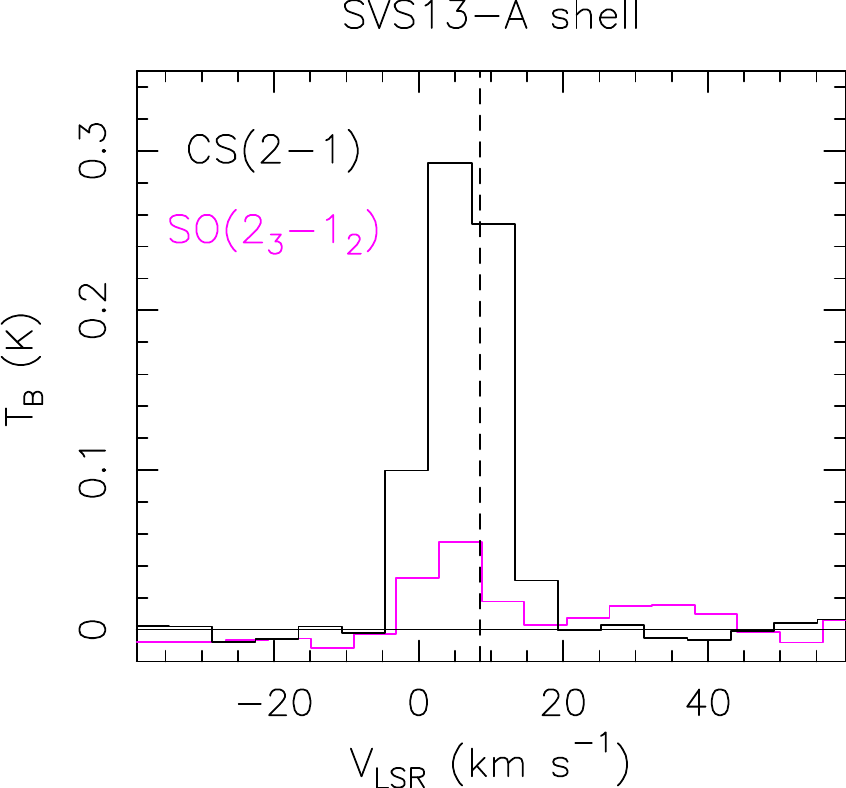}
\caption{Line profiles in CS(2--1), and SO(2$_3$--1$_2$), in brightness temperature scale, towards the peak of the CS emission
in the SVS13-A southern molecular shell:
$\alpha(\rm {J2000})$ = 03$^{h}$ 29$^{m}$ 05$\fs$317, $\delta(\rm {J2000})$ = +31$^{\degr}$ 15$^{\arcmin}$ 30$\farcs$09.
The black vertical line at +8.5 km s$^{-1}$
is for the systemic emissions. The CS lines is brighter than the SO one, although a word of caution is needed given the position lies outside the FoV of the images.}
\label{fig:spectrashell}
\end{center}
\end{figure}

Figure \ref{fig:spectrashell} shows the CS(2--1), and SO(2$_3$--1$_2$) spectra extracted at the CS emission peak 
in the southern molecular shell (Figs. \ref{fig:mom0}
and \ref{fig:sofia}).

\section{CH$_3$OH emission in the SVS13-B jet}
\label{sec:spectralfit}

\begin{figure}
\begin{center}
\includegraphics[angle=0,scale=0.3]{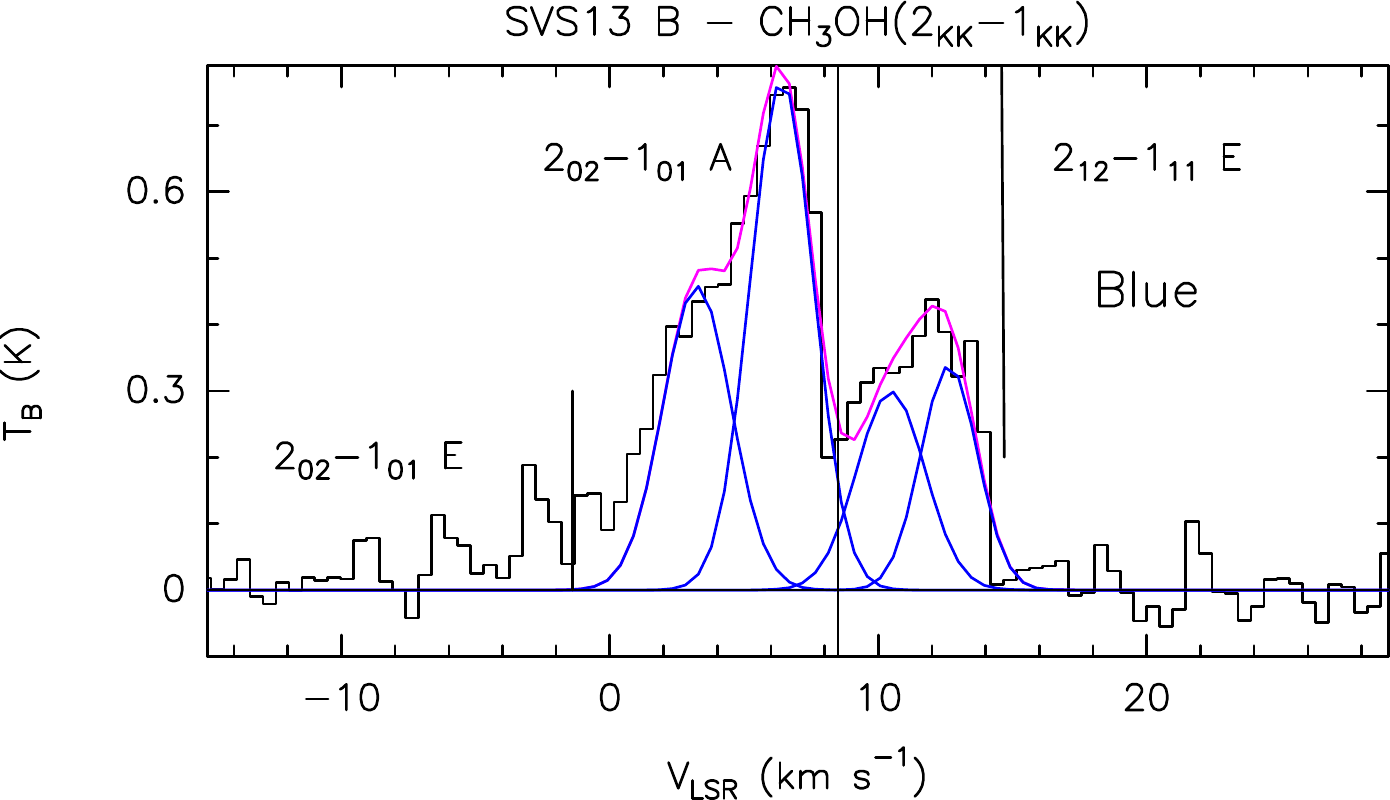}
\caption{CH$_3$OH(2$_{\rm kk}$--1$_{\rm kk}$) spectrum (in  brightness temperature, $T_{\rm B}$, scale) extracted at the blue-shifted emission peak position (see Sect. \ref{sec:jetB}). 
The transitions producing the methanol profile are labelled, with the corresponding frequencies  marked by vertical black segments (see Table \ref{tab:lines}). The CH$_3$OH spectrum is centred at the frequency of the 2$_{\rm 0,2}$--1$_{\rm 0,1}$ A transition: 96741.38 MHz.
The black vertical line at +8.5 km s$^{-1}$ is for the systemic velocity \citep{Podio2021}. The 2$_{\rm 0,2}$--1$_{\rm 0,1}$ A and 2$_{\rm 1,2}$--1$_{\rm 1,1}$ E lines have been fit (magenta contour) by assuming for each line two Gaussian components with a FWHM of 3 km s$^{-1}$  shifted by  --2.0 km s$^{-1}$, and --5.3 km s$^{-1}$, respectively (in blue).} 
\label{fig:methanol-spectralfit}
\end{center}
\end{figure}

Figure \ref{fig:methanol-spectralfit} shows the CH$_3$OH(2$_{\rm kk}$--1$_{\rm kk}$) emission pattern observed towards the blue-shifted emission peak position (see Sect. \ref{sec:jetB}). The spectrum is centred at the frequency of the 2$_{\rm 0,2}$--1$_{\rm 0,1}$ A transition (96741.38 MHz). The modeled spectral pattern (in magenta) is obtained by postulating that each spectral line is composed of two Gaussian components with a FWHM of 3 km s$^{-1}$ and centered at different velocities: one shifted by --2.0 km s$^{-1}$ and another one by --5.3 km s$^{-1}$ (in blue).

We report here also the images of the molecular jet driven by the SVS13-B
using the channel maps of the CH$_3$OH red- and blue-shifted emission. 
As the methanol emission is due to the blending of the different components, the figure shows the red-shifted emission of the 2$_{\rm 1,2}$--1$_{\rm 1,1}$ E (Upper panels), and the blue-shifted emission of the 2$_{\rm 1,2}$--1$_{\rm 1,1}$ A (Lower panels).

\begin{figure*}
\begin{center}
\includegraphics[angle=0,scale=0.46]{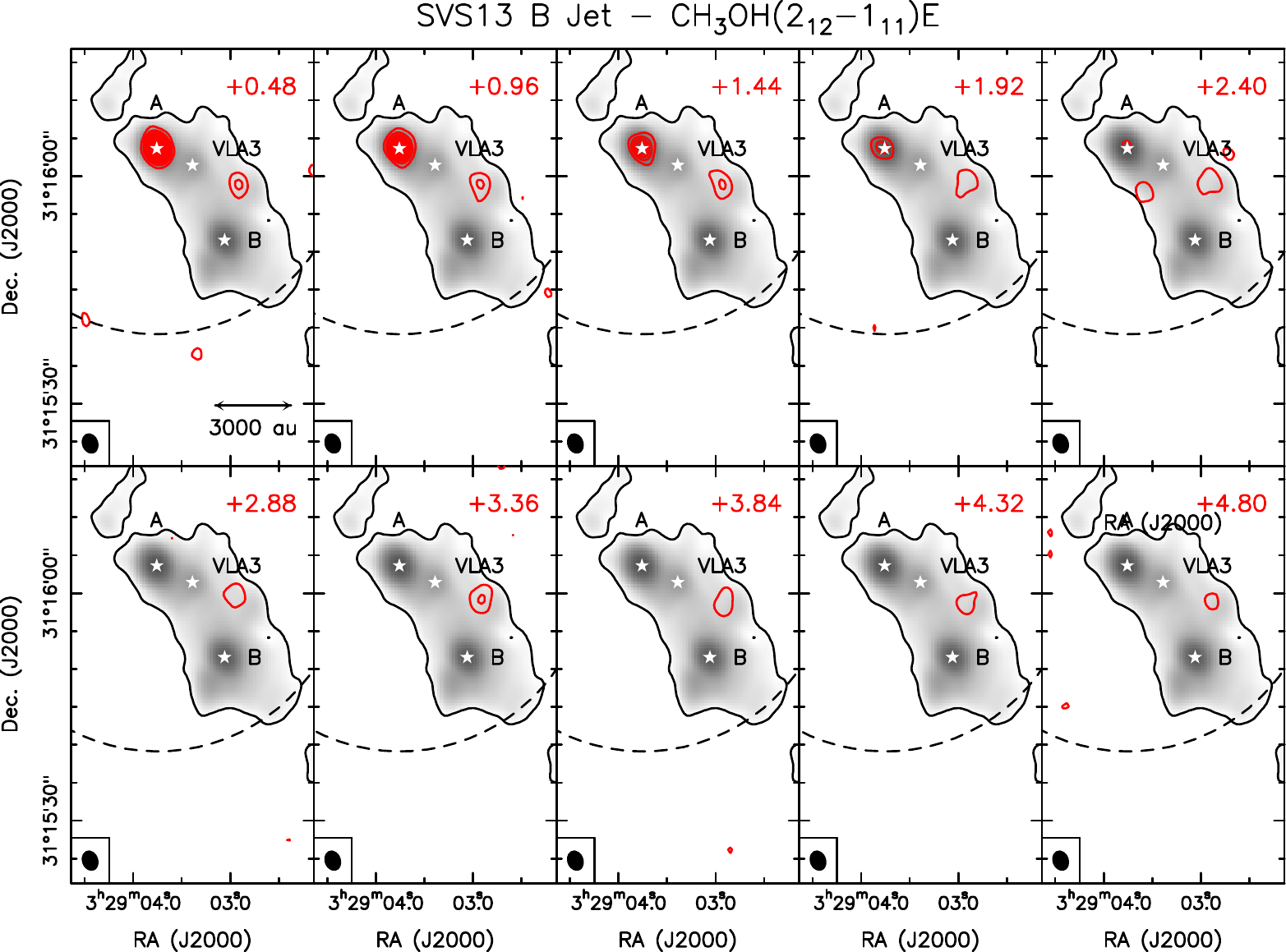}
\includegraphics[angle=0,scale=0.65]{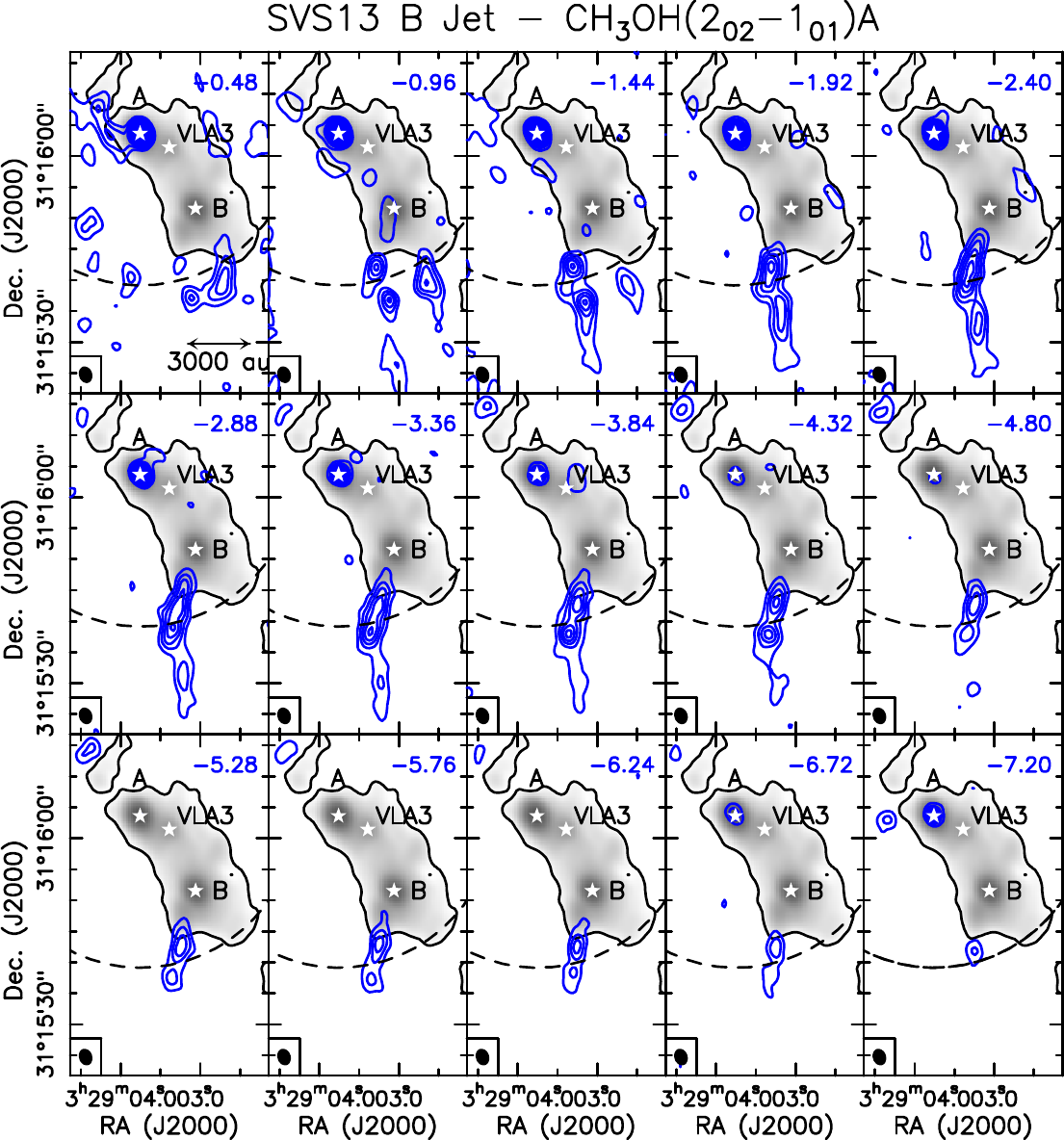}
\caption{Channel maps of the CH$_3$OH red- and blue-shifted emission. 
Each panel shows the emission shifted in velocity with respect to the systemic velocity \citep[+8.5 km s$^{-1}$, ][]{Podio2021} by the value given in the upper right corner. 
As the methanol emission is due to the blending of the different components, the figure shows the red-shifted emission of the 2$_{\rm 1,2}$--1$_{\rm 1,1}$ E (upper panels), and the blue-shifted emission of the 2$_{\rm 1,2}$--1$_{\rm 1,1}$ A (lower panels).
In grey scale (and black contour) the 3mm continuum image (Fig. \ref{fig:continuum}) is shown. First contours and steps are 3$\sigma$ (6 mJy beam$^{-1}$).
The positions of the A, VLA3, and B continuum peaks are marked by  white  stars (Table \ref{tab:continuum}). The synthesised beams are shown in the bottom-left, while the dashed circles delimitate the FoV
(Tab. \ref{tab:lines}).
}
\label{fig:ch3oh-channels}
\end{center}
\end{figure*}

\section{Comparison between $J$ = 2--1 and $J$ = 5--4 SiO emission from the SVS13-B jet}

\begin{figure}
\begin{center}
\includegraphics[angle=0,scale=0.43]{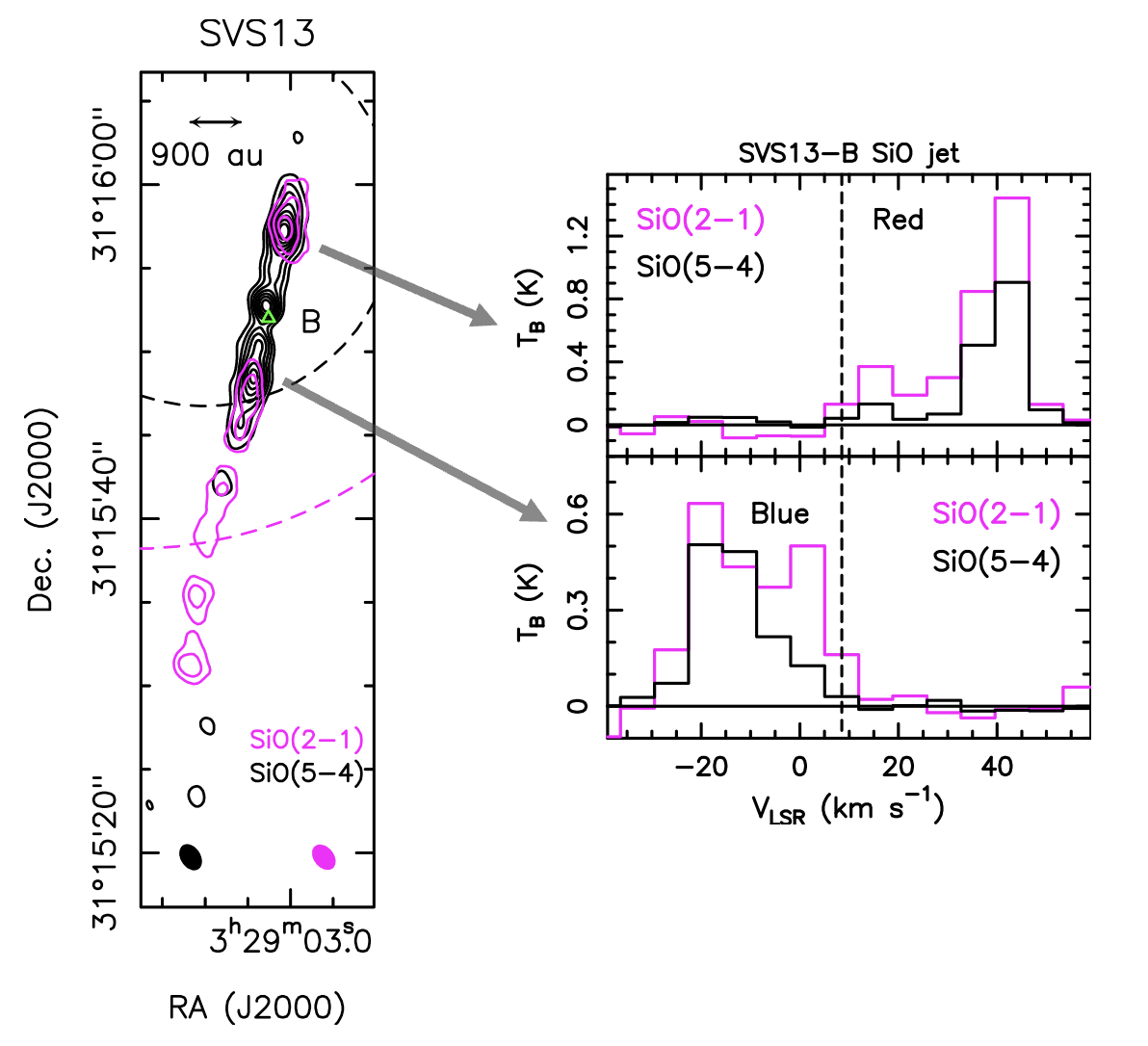}
\caption{Left: The red- and blue-shifted bipolar SVS13-B SiO jet as traced by the
$J$ = 2--1 (magenta, this paper), $J$ = 5--4 \citep[black, from][]{Podio2021}. The SiO(5--4) image by \citet{Podio2021} has been smoothed to the spatial resolution of the present SiO(2--1) map
(1$\farcs$72 $\times$ 1$\farcs$20, 39$\degr$). 
First contours and steps are 3$\sigma$ (96 mJy km s$^{-1}$ beam$^{-1}$ for $J$ = 2--1, and 75 mJy km s$^{-1}$ beam$^{-1}$ for $J$ = 5--4), and 2$\sigma$, respectively. The synthesised beams are shown in the bottom-left corners. 
The dashed circles delimitate the FoV of the images. The position of the B continuum peak is marked by a green triangle (Table \ref{tab:continuum}).
Right: The SVS13-B jet:  SiO(2--1) and SiO(5--4) line profiles, in brightness temperature scale, towards the SiO red- (Upper panel), and blue-shifted (Lower panel) emission peaks. The positions where the spectra have been extracetd are marked, in Fig. \ref{fig:comparison}, by blue and red crosses. For SiO(2--1), the dataset with the beam of 1$\farcs$72 $\times$ 1$\farcs$20 has been used for a proper comparison 
with the SiO(5--4) data by \citet{Podio2021}.
The black vertical line at +8.5 km s$^{-1}$
is for the systemic velocity.} \label{fig:calypso-composed}
\end{center}
\end{figure}

Figure \ref{fig:calypso-composed} compares the intensity of the J = 2--1 line with that of SiO(5--4) reported, in the context of the CALYPSO IRAM PdBI Large Program, by \citet{Podio2021}. The SiO(5--4) image by \citet{Podio2021} has been smoothed to the spatial resolution of the present SiO(2--1) map (1$\farcs$72 $\times$ 1$\farcs$20, 39$\degr$) for a proper comparison and consequently to constrain the physical conditions of the SVS13-B jet.

\section{Channel maps of SiO and SO emission from the jet driven by SVS13-A}
\label{sec:outflowA}

\begin{figure*}
\begin{center}
\includegraphics[angle=0,scale=0.5]{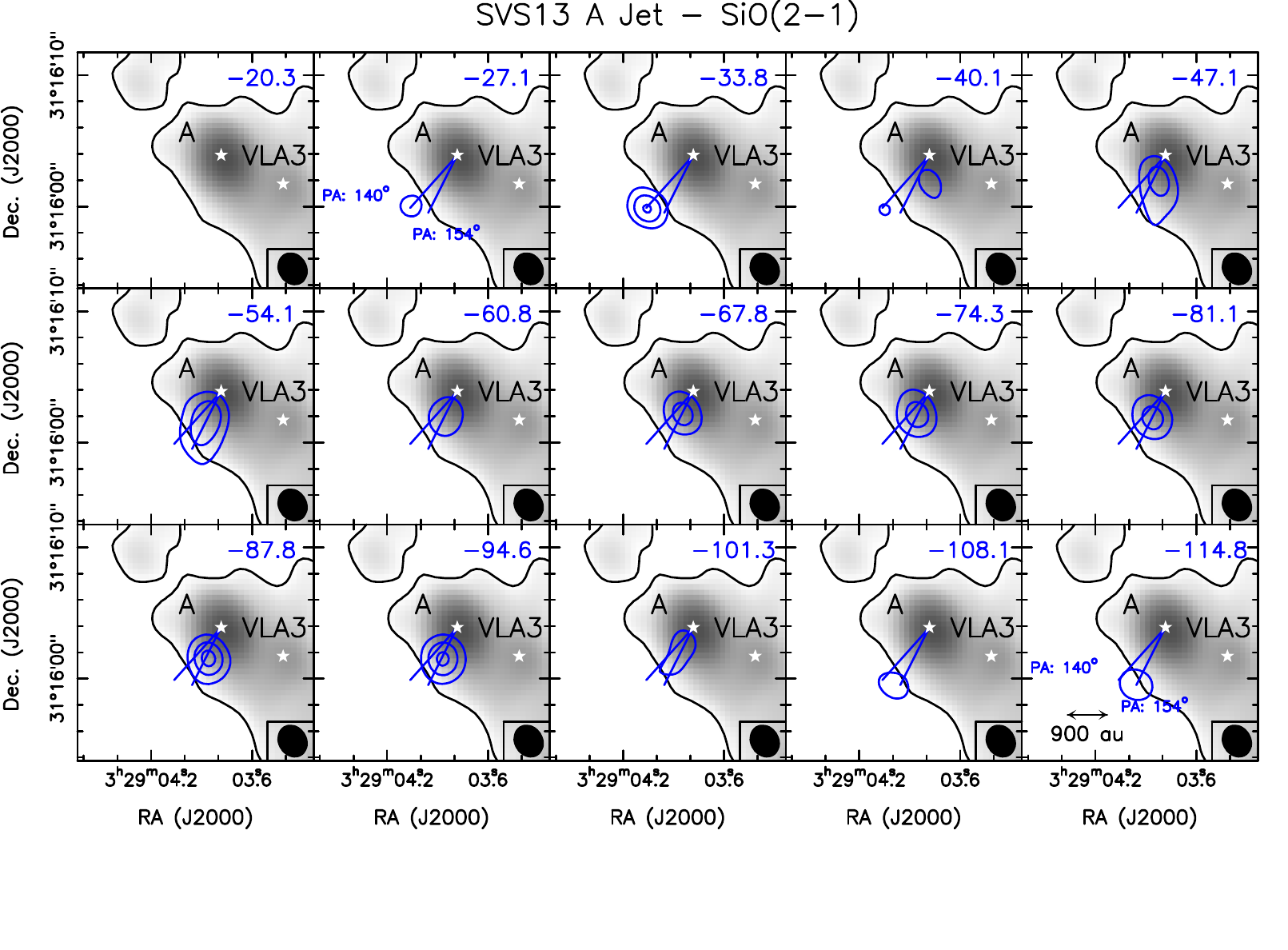}
\caption{Channel maps of the SiO(2--1) blue-shifted emission towards SVS13-A. Each panel shows the emission shifted in velocity with respect to the systemic velocity  \citep[+8.5 km s$^{-1}$, ][]{Podio2021} by the value given in the upper right corner. The positions of SVS13-A and
VLA3 continuum peaks are marked by  white stars, the beam is in the bottom-right corner. In grey scale (and black contour) the 3mm continuum image  is shown (Fig. \ref{fig:continuum}).
First contours and steps are 3$\sigma$ (24 mJy beam$^{-1}$).
The pair of blue lines delineates the jet's trajectory, as marked by the slowest ($\sim-27$ km s$^{-1}$, PA $\sim 140\degr$) and the fastest ($\sim - 115$ km s$^{-1}$, PA $\sim 154\degr$) emissions.}
\label{fig:sio-channelsA}
\end{center}
\end{figure*}

\begin{figure*}
\begin{center}
\includegraphics[angle=0,scale=0.5]{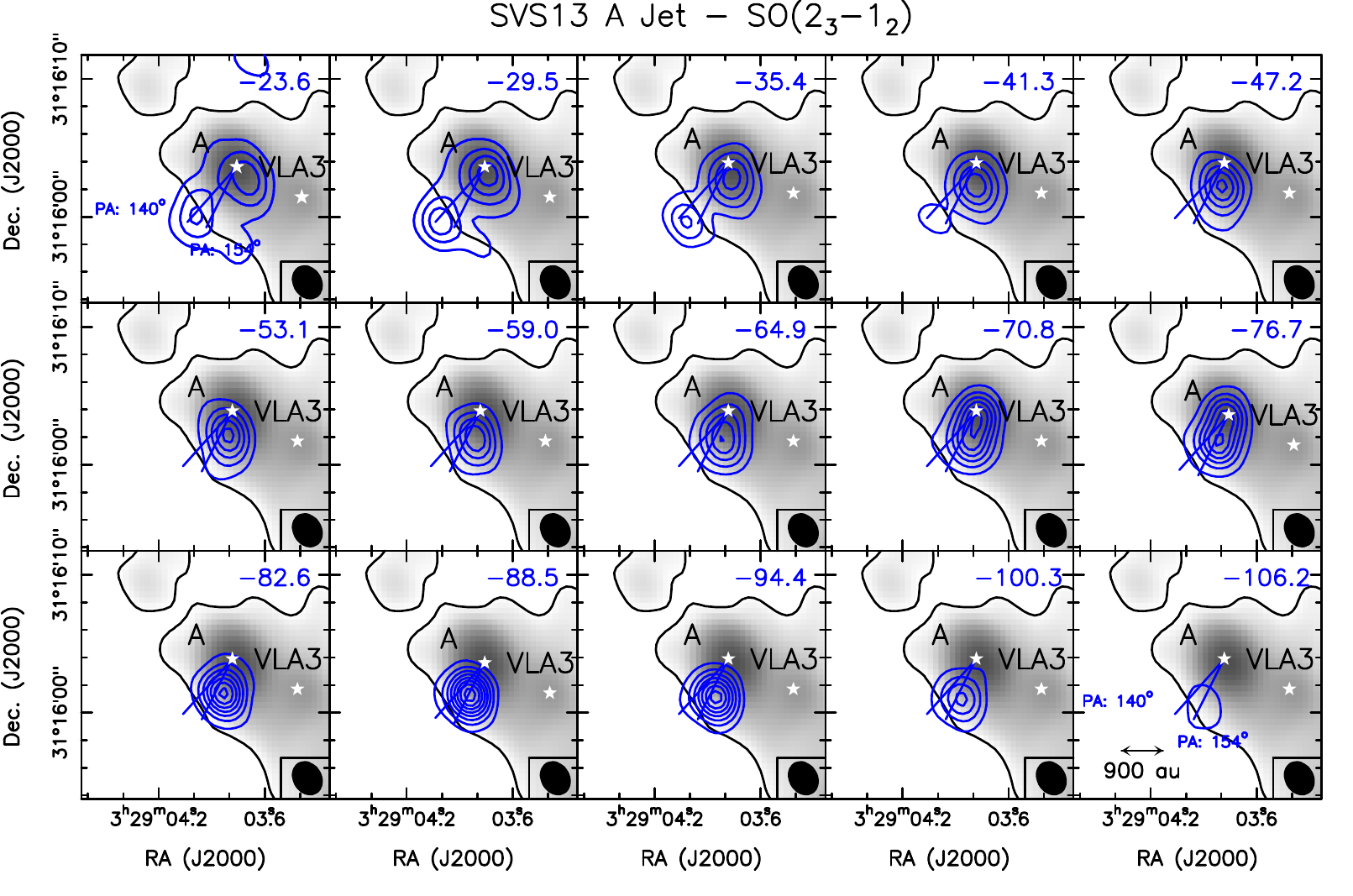}
\caption{Channel maps of the SO(2$_3$--1$_2$) blue-shifted emission towards SVS13-A. Each panel shows the emission shifted in velocity with respect to the systemic velocity  \citep[+8.5 km s$^{-1}$, ][]{Podio2021} by the value given in the upper right corner. The positions of SVS13-A and
VLA3 continuum peaks are marked by  white stars, the beam is in the bottom-right corner. In grey scale (and black contour) the 3mm continuum image  is shown (Fig. \ref{fig:continuum}).
First contours and steps are 3$\sigma$ (21 mJy beam$^{-1}$).
The pair of blue lines delineates the jet's trajectory, as marked by the slowest ($\sim -24$ km s$^{-1}$, PA = $\sim 140\degr$) and the fastest ($\sim - 106$ km s$^{-1}$, PA $\sim154\degr$) velocities.}
\label{fig:so-channelsA}
\end{center}
\end{figure*}

We report here the images of the molecular jet driven by the SVS13-A 
binary system.
The channel maps of SiO(2--1) and SO(2$_3$--1$_2$) emissions are shown
in Figs. \ref{fig:sio-channelsA} and \ref{fig:so-channelsA}, respectively.
The SiO(2--1) emission consists of spatially unresolved ($\leq$ 450 au) clumps. By referencing the location of the SVS13-A binary system (here spatially unresolved), we can observe that clumps moving at various speeds trace out different Position Angles (PA). 
The slowest moving clump  (approximately --27 km/s relative to $V_{\rm sys}$) is aligned at a PA of roughly 140$\degr$, whereas the fastest moving clump ($\simeq$ 115 km/s) aligns at a PA of about 154$\degr$.
While SO channel maps display velocities that are slower compared to SiO (reaching speeds up to about --106 km s$^{-1}$ with respect to systemic), SO(2$_3$--1$_2$) channel maps exhibit a comparable kinematics.
These results are in agreement with the analysis by \citet{Lefevre2017}, based on IRAM-PdBI SiO(5--4) maps, who depicted a wiggling jet near the SVS13-A location. 
This jet has PA$\sim 150\degr \pm 10\degr$, aligning with
the CO jet reported by \citet{Bachiller2000}. The SiO and SO emission probed by our observations does not trace the ejection process responsible for the HH7--11 chains, which are placed along a PA of about 
130$\degr$ \citep[see e.g. Fig. 5 by][and references therein]{Lefevre2017}.

\end{appendix}

\end{document}